\documentclass[pdflatex,sn-mathphys-num]{sn-jnl}

\usepackage{graphicx}
\usepackage{amsmath}
\usepackage{amssymb}
\usepackage{amsfonts}
\usepackage{booktabs}

\usepackage{amscd}
\usepackage[normalem]{ulem}
\usepackage{float}
\usepackage{subcaption}
\usepackage{bbm}
\usepackage{mathtools}
\usepackage{tikz}
\usepackage{tikz-3dplot}
\usepackage{quiver}
\usetikzlibrary{arrows.meta,decorations.pathmorphing,positioning,fit,shapes.geometric,shapes.misc}
\usetikzlibrary{calc}
\theoremstyle{thmstyleone}
\newtheorem{theo}{\bfseries Theorem}[section]
\newtheorem{prop}[theo]{\bfseries Proposition}

\theoremstyle{thmstylethree}
\newtheorem{rem}[theo]{\bfseries Remark}
\newtheorem{defn}[theo]{\bfseries Definition}
\theoremstyle{thmstyleone}

\newcommand{\mbb}[1]{\mathbb{#1}}

\newcommand{\lrb}[1]{\left\{#1\right\}}
\newcommand{\lrs}[1]{\left[#1\right]}
\newcommand{\lrp}[1]{\left(#1\right)}

\newcommand{\C}[1]{\mathbb{C}}
\newcommand{\CP}{\mathbb{C}\mathrm{P}}

\newcommand{\ket}[1]{\lvert #1 \rangle}

\DeclarePairedDelimiterX{\inp}[2]{\langle}{\rangle}{#1, #2}
\newcommand{\Trp}[0]{\text{T}}

\begin{document}

\title{Visualizing the state space and transformations of higher order quantum logics via toric geometry}

\author*[1]{\fnm{Steven} \sur{Bleiler}\,\orcid{https://orcid.org/0009-0008-1871-194X}}\email{bleilers@pdx.edu}
\author[2]{\fnm{Shanyan} \sur{Chen}\,\orcid{https://orcid.org/0009-0005-0837-5215}}\email{sychen@cnu.edu.cn}
\author[1]{\fnm{Emma L.} \sur{O'Neil}}\email{emmao@pdx.edu}
\equalcont{These authors contributed equally to this work.}
\author[1]{\fnm{J. Eliot} \sur{Reich}}\email{reichel@pdx.edu}
\equalcont{These authors contributed equally to this work.}
\author[1]{\fnm{Julia} \sur{Rezvani}}\email{rezvani@pdx.edu}
\equalcont{These authors contributed equally to this work.}
\author[1]{\fnm{Elijah} \sur{Whitham-Powell}}\email{ew4@pdx.edu}
\equalcont{These authors contributed equally to this work.}
\author[3]{\fnm{Ali} \sur{Al-Bayaty}\,\orcid{https://orcid.org/0000-0003-2719-0759}}\email{albayaty@pdx.edu}
\equalcont{These authors contributed equally to this work.}
\author[3]{\fnm{Jerzy} \sur{Jegier}}\email{jerzyjeg@gmail.com}
\author[3]{\fnm{Sonia} \sur{Yang}}\email{slysonia10@gmail.com}
\author[1]{\fnm{Marek} \sur{Perkowski}\,\orcid{https://orcid.org/0000-0002-0358-1176}}\email{h8mp@pdx.edu}

\affil*[1]{\orgname{Portland State University}, \orgaddress{\city{Portland}, \state{Oregon}, \country{USA}}}
\affil[2]{\orgname{Capital Normal University}, \orgaddress{\city{Beijing}, \country{China}}}
\affil[3]{\orgname{Independent Researcher}}

\abstract{We propose some new uses of toric variety structures in the study of quantum computation for small radices. In particular, we observe the concurrence of the equivalence classes of quantum states under quantum measurement and the orbits of the toric geometric structure of the state space.  Visualizations of these state spaces and of certain fundamental unitary transformations in binary and ternary quantum logic and a method to develop new transformations based on these visualization techniques are presented. Transformations discussed included minimal universal sets for permutative ternary quantum circuits. In addition, general structures and synthesis methods based on quantum multiplexers are presented.  A general framework for the design of optimal ternary quantum transformations and circuits is additionally presented. Finally, a number of open research areas that are extensions of the work presented herein are given.}

\keywords{quantum trits, qudits, qubits, toric geometry, visualization}

\maketitle

\noindent\textbf{Corresponding Author}: Steven Bleiler

\section*{Introduction}

For some years now there has been a clear demand expressed in the engineering literature for useful and natural geometric representations of the analogue of the Bloch sphere for individual qubits for pairs of qubits and for order three logic, i.e.\ for the joint state space of a pair of qubits and of the state space of the quantum trit.  Surprisingly absent from the engineering literature, such geometric representations have been used and exploited by mathematicians studying the mathematical properties of the complex projective spaces (and generalizations thereof) for some 75 years \cite{ewald1996}, even though it did take nearly 50 years for some of these representations to appear in the physics literature, see, for example \cite{bengtsson2002}.  In \cite{bengtsson2002}, the toric variety structure (also known in the mathematics literature as the toric geometry) of the finite dimensional complex projective spaces is used by the authors to illustrate geometrically various quantum phenomena such as separability and entanglement for a pair of qubits.

Recent advances in qubit logic synthesis based on the geometry of the Bloch sphere \cite{albayaty2024bsa}, \cite{albayaty2023galangenericarchitecturelayoutaware}, \cite{albayaty2024cala} suggest that having such representations will lead to better transformation design and logic synthesis for higher radix quantum computing through the sole use of the ``native'' transformations of a given implementation for quantum computation. The rapid advent of topological quantum computation with its natural order three logic has added a fresh urgency for these geometric representations to become better known and exploited by the engineering community.  Our research project's purpose is to present in a straightforward way the toric variety structures of the low dimensional complex projective spaces in a manner accessible to engineers and engineering students and to indicate a few of the advances of understanding possible through their use, giving particular attention to the visualization of unitary transformations important for quantum computation.  In this way, we hope to expand the interactions between the mathematical and engineering communities.

In particular, following our previous work and that of other authors \cite{goss2022high,lawrence2004mutually,moraga2014some,pudda2024generalised}, we illustrate the use of the toric variety visualization of the state space of a quantum trit and unitary transformations of this space to create efficient versions of general ternary quantum circuits. 

The paper begins by establishing the necessary mathematical background regarding group actions, the tori, toric groups, and toric geometry. With these topics in hand, the paper proceeds to express a toric geometry visualization of the state spaces of the fundamental logical units of quantum computation. This visualization is then extended to the physical transformations of these logical units and concludes with applications to the design and synthesis of quantum circuits realizing various quantum transformations and quantum algorithms.

Scientific contributions include the observation of the coincidence of toric geometric structures and the equivalence classes of states under quantum measurement and a technique for visualizing the state spaces and unitary transformations of individual quantum logic elements. This visualization technique then allows for the development of novel factorizations of quantum transformations. These factorizations allow for the more efficient synthesis of quantum circuits realizing various quantum algorithms. The visualization technique is perfectly general, and can be employed in several other contexts.

\section{Mathematical background}
\subsection{Tori and toric geometry}

There is significant interest in the visualization of the state space of the quantum trit, as evidenced by the number of prospective visualizations appearing in the literature.
Many of these visualizations and parameterizations \cite{mosseri2001} are based on the cellular structure of $\CP^n$, as appearing in \cite{hatcher2000}. Both these visualizations and the visualization presented here rely on normalization of the homogeneous coordinates of a pure quantum state.
Other visualization techniques leverage operator bases to parametrize the state space of an arbitrary mixed quantum computation logical unit, frequently resulting in parameterized state spaces which contain representations of nonphysical quantum states which must be worked around \cite{kurzynski2016} \cite{eltschka2021} \cite{sharma2024}. When considering only the pure state of a qudit, these concerns are simplified.
The visualization presented here is thoroughly grounded in the theory of toric varieties as originally developed by F. Hirzebruch \cite{hirzebruch1950}. It has the advantage of providing a unified visualization in arbitrary dimensions which does not admit representations of nonphysical states, enabled by a decomposition of states by ``internal'' phases.

Our exposition will employ mathematical objects which may not be familiar to many engineering professionals.

\begin{defn}
The \emph{$n$-tori} $T^n$ are topological spaces which are diffeomorphic to the cartesian product of $n$ copies of the unit circle $S^1$.
\end{defn}
These topological spaces admit a much richer algebraic structure from the theory of Lie groups as the unit circle $S^1$ is the underlying $C^\infty$ manifold of the Lie group of unit complex numbers, also known as the group of $1 \times 1$ unitary matrices, $U(1)$.  Considered in this way the various $n$-tori (including the ``degenerate'' tori $T^0$, consisting of a single point, and $T^1$, consisting of a single circle) are all compact abelian Lie groups.

\begin{defn}
    A \emph{group action} of a group $G$ on a set $A$ is a map from $G \times A$ to $A$ (written as $ga$ for all $g \in G$ and $a \in A$) satisfying the following properties:
    \begin{itemize}
        \item  $g_1(g_2 a) = (g_1 g_2)a$, for all $g_1, g_2 \in G$, $a \in A$.
        \item  $1a = a$, for all $a \in A$.
    \end{itemize} 

    If $G$ is a Lie Group (smooth) and $A$ is a complex manifold, then the group action is called a \emph{smooth complex action} \cite{dummit2004}.
\end{defn}

\begin{defn}
    Let $G$ be a group acting on a nonempty set A. The equivalence class \{$ga \mid g \in G$ \} is called the \emph{orbit} of $G$ containing $a$ \cite{dummit2004}.
\end{defn}

\begin{defn}
Complex projective $n$-space $\CP^n$ is defined as the quotient of the $n{+}1$ dimensional affine complex vector space less the zero vector, $\mbb{C}^{n+1}\setminus\{\vec{0}\}$, by the non-zero complex numbers, $\mbb{C}\setminus\{0\}$, by setting $\lambda \mathbf{v} \equiv \mathbf{v}$ whenever $\lambda \in \mbb{C}\setminus\{0\}$. 
\end{defn}
In the language of group actions, complex projective $n$-space is the orbit space of the action of the multiplicative group of nonzero complex numbers on the set $\mathbb{C}^{n+1}\setminus\{\vec{0}\}$ given by scalar multiplication.

This is precisely what physicists mean by phase equivalence, and the axioms of quantum mechanics stipulate that the state spaces of quantum systems form precisely such a space, though in the general quantum mechanical case, possibly infinite dimensional.

For visualization and linear algebraic purposes, it is useful to note that the length of a given non-zero vector can be regarded as just a real valued phase and that we can express $\mbb{C}^{n+1}\setminus\{\vec{0}\}$ as the cartesian product of the unit sphere $S^{2n+1}$ and the real interval $(0,\infty)$ and similarly express the non-zero complex numbers as the cartesian product of the unit complex numbers $U(1)$ and the real interval $(0,\infty)$.

Quotienting cancels the cartesian product with $(0,\infty)$, and we see $\CP^n$ expressed as the quotient of $S^{2n+1}$ by the scalar multiplication action of $U(1)$, i.e.\ we regard $\lambda \mathbf{v} \equiv \mathbf{v}$ for all $\lambda \in U(1)$.  For $n = 1$, this quotient function is the (right hand) Hopf map. Expressing the 3-sphere in complex affine coordinates $(z_0, z_1)$ in $\mbb{C}^2$ with $|z_0|^2 + |z_1|^2 = 1$, we have the Hopf map explicitly expressed as:
\[
    S^3 \to \CP^1 : (z_0, z_1) \mapsto z_1^{-1} z_0 \in \mbb{C} \cup \{1/0\} \cong S^2.
\]

\begin{figure}[H]
    \centering
    \includegraphics[width=0.75\linewidth]{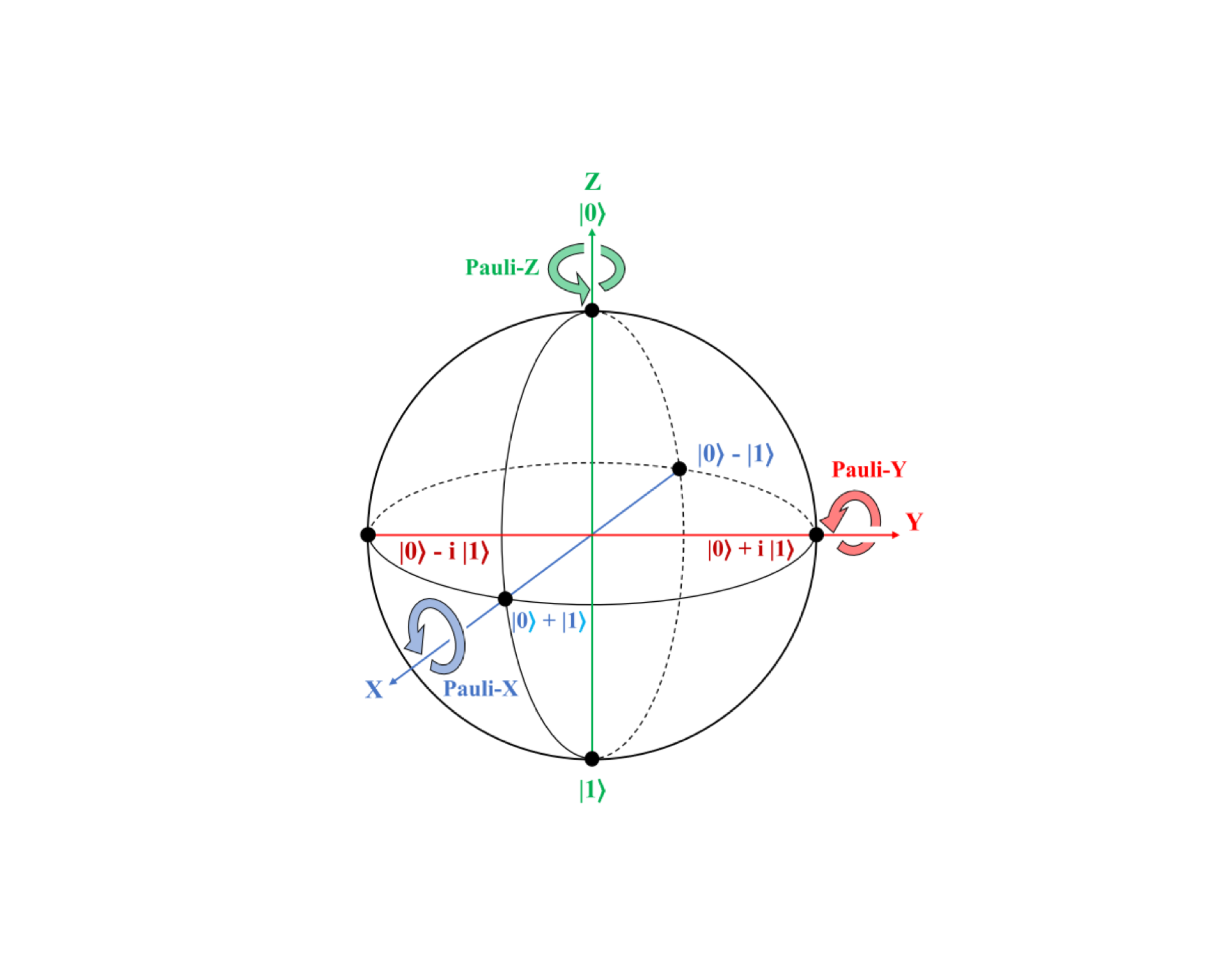}
    \caption{A standard picture of the Bloch sphere $\CP^1$.}
    \label{fig:bloch}
\end{figure}

Shown in Figure~\ref{fig:bloch}, $\CP^1$ represents the state space of a quantum bit.  Using it we can visualize the action of various single qubit transformations such as the three Pauli $\pi$-rotations (around the X, Y, and Z axis respectively).  Engineers and physicists use this visualization to design not just new transformations, but also new factorizations of existing transformations, transformations which in a given implementation of quantum computation may be extremely expensive to realize, while their new factors are not.

\begin{figure}[H]
    \centering
    \includegraphics[width=0.35\linewidth]{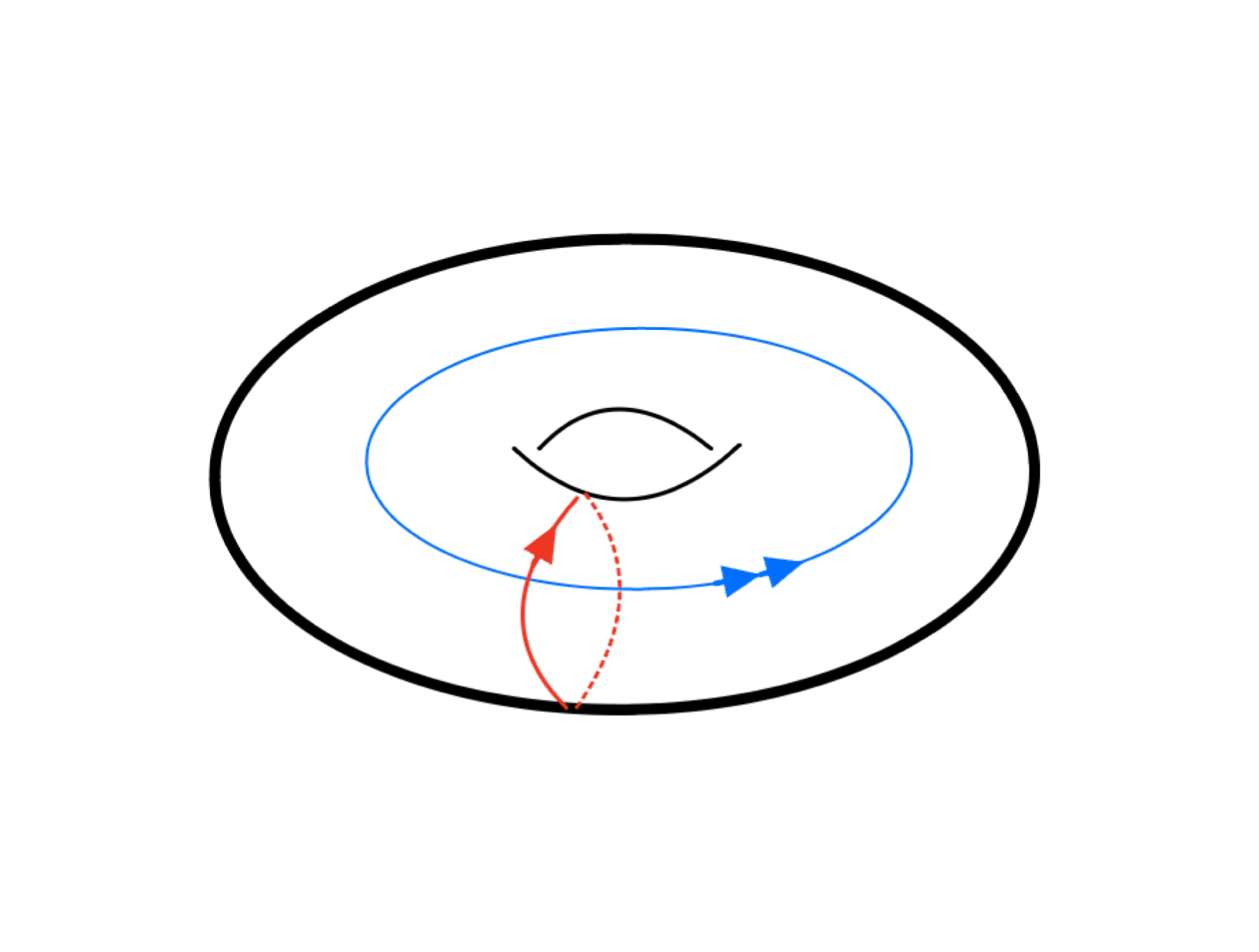}
    \caption{A 2-torus $T^2$ in Euclidean 3-space.}
    \label{fig:torus}
\end{figure}

\begin{defn}
Complex manifolds are said to admit a \emph{toric geometry} when they admit a smooth complex action of an $n$-torus. The expression of a complex manifold that admits a toric geometry in terms of the space of orbits of the action and the individual orbits themselves, which are geometric tori of dimension $\leq n $ is known as a \emph{toric variety}\cite{ewald1996}.
\end{defn}
Of interest in the study of toric geometry is the geometry of the space of orbits of the action, in addition to the geometric structure of the individual orbits themselves.
Recall that a geometric torus of dimension $n \geq 2$ does not isometrically embed in Euclidean 3-space as each individual $U(1)$ orbit in a given factor must have the same length.  Contrast this with the 2-torus illustrated in Figure~\ref{fig:torus}, where the meridional circles (i.e.\ the ones around the ``arm'' of the torus) all do have the same length, but the longitudinal circles (i.e.\ the ones around the ``hole'' of the torus) do not.

\begin{figure}[H]
    \centering
    \includegraphics[width=0.5\linewidth]{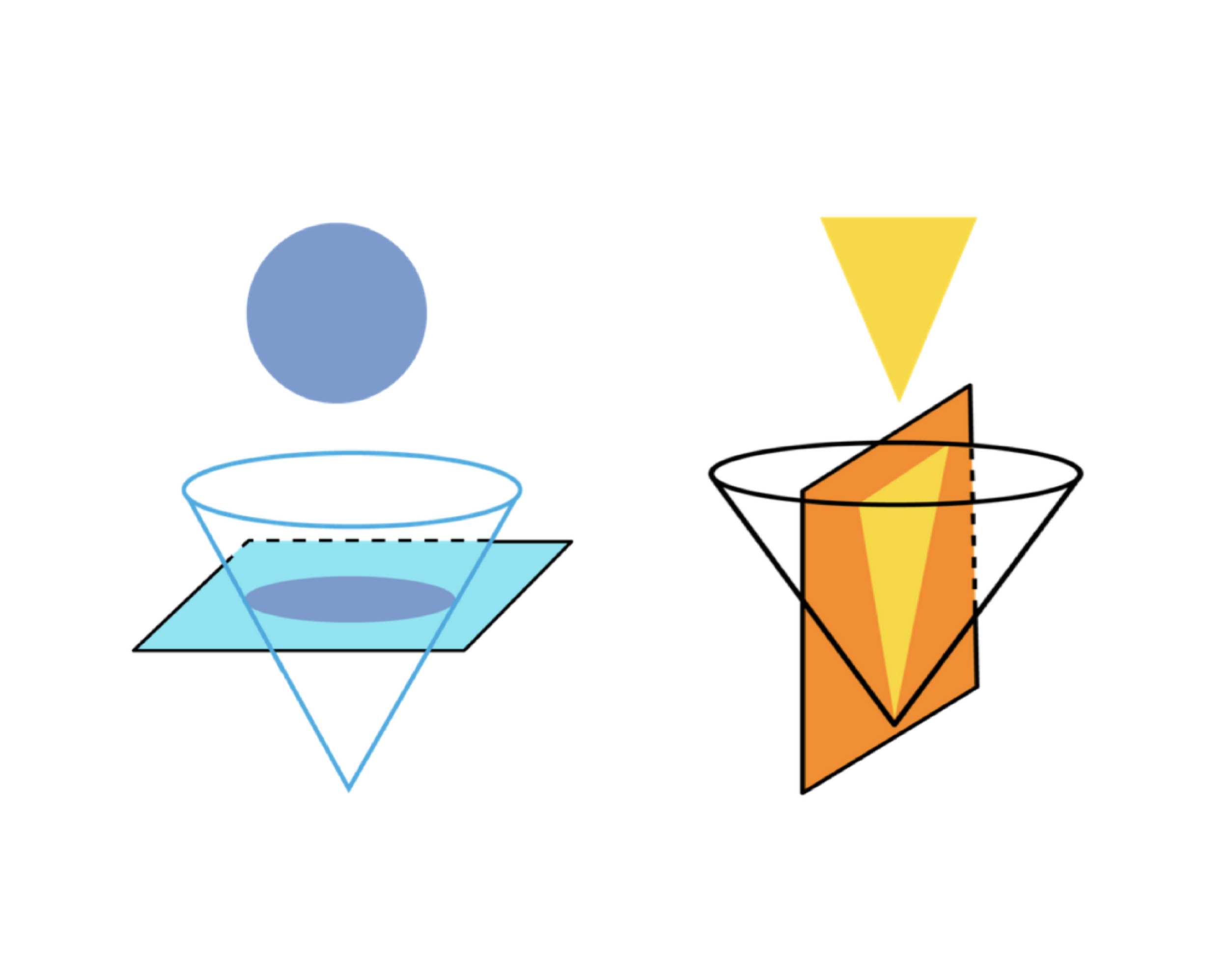}
    \caption{A 2-dimensional visualization of a solid cone.}
    \label{fig:cone}
\end{figure}

The structures in toric geometry are typically expressed therefore as a pair of transverse factors, the space of orbits, which in the cases we will study are simply the standard $n$-simplices of real convex linear combinations lying in the non-negative hyperoctants in Euclidean $n{+}1$ dimensional space, and the individual tori lying over the various points of this parametrizing space, much like the way mathematicians might express a solid cone to a 2-dimensional observer as meeting different perpendicular planes in the very different geometric forms of a circle or a triangle, as indicated in Figure~\ref{fig:cone}.

\subsection{Visualizing complex projective spaces, quantum measurement and its relation to toric geometry}

\begin{defn}
     A \emph{qubit} (short for \emph{quantum bit}) is the basic unit of information in quantum computing, analogous to a classical bit in classical computing. Qubits exist in states, and while $\ket{0}$ and $\ket{1}$ are possible states for a qubit, it is also possible to form linear combinations of states, called superpositions in the physics literature.
     \[\ket{\psi} = \alpha\ket{0} + \beta\ket{1}\]
     The numbers $\alpha$ and $\beta$ are complex numbers, and the basis states $\ket{0}$ and $\ket{1}$ form an orthonormal basis for the Bloch sphere $\mathbb{C}P^1$, known as the computational basis \cite{chuang2010}.
\end{defn}

Other quantum systems employ higher order logics. For example, a \emph{qutrit} is analogous to a qubit, but with three computational basis states $\ket{0}$, $\ket{1}$, and $\ket{2}$, a \emph{ququadit} has four computational basis states $\ket{0}$, $\ket{1}$, $\ket{2}$, and $\ket{3}$, and a general \emph{qudit} has $d$ computational basis states $\ket{0}, \dots, \ket{d-1}$.

There is a strong correlation between the maximal subsets of $\CP^n$ where each state in the subset quantum measures identically, that is, the probability of observing a given basis element is the same for each element in the set, and the natural toric geometry structure on $\CP^n$.

This is easily seen in the Bloch sphere as the decomposition into the set of latitudinal circles unioned with the set consisting of the two poles.  In coordinates, up to global phase, every homogeneous coordinate $(z_0, z_1)$ not the ``pole'' $(0,1)$ is phase equivalent to a coordinate of form $(x_0, \lambda x_1)$ with the $x_i$ non-negative real numbers, $\lambda \in U(1)$, and such that $x_0^2 + x_1^2 = 1$.  This decomposition is precisely the set of orbits of the toric action of $U(1)$ on $\CP^1$ given by $\lambda \cdot (z_0, z_1) = (z_0, \lambda z_1)$. \begin{prop}
The decomposition of $\CP^1$ into latitudinal circles and points is precisely the decomposition of $\CP^1$ given by declaring states to be equivalent when they behave identically under quantum measurement.    
\end{prop}
\begin{proof}
    Distinct quantum states with homogeneous states $(z_0, z_1), (z_0^\prime, z_1^\prime)$ measure identically in the $|0\rangle,|1\rangle$ basis if and only if they induce the same observation probabilities $|z_0|^2$ and $|z_1|^2$, over the basis states $\ket{0}$ and $\ket{1}$ respectively. Equivalently, the coordinate moduli $|z_0|=|z_0^\prime| = x_0$ and $|z_1|=|z_1^\prime| = x_1$. Then up to common global phase, $(z_0,z_1) \equiv (x_0, \lambda x_1)$ and $(z_0^\prime, z_1^\prime) \equiv (x_0, \lambda^\prime x_1)$ for unit complex numbers $\lambda$ and $\lambda^\prime$. This is exactly the condition for $(z_0, z_1)$ and $(z_0^\prime, z_1^\prime)$ to lie in the same toric orbit.
\end{proof}

This coordinate formula for the toric geometry on $\CP^1$ also expresses the described visualization of toric geometry as an orbit space plus a set of orbits.  Each orbit under the torus action can be uniquely represented by a real number pair $(x_0, x_1)$ of coordinate lengths, and with the two polar exceptions, each orbit represented by a copy of $U(1)$.  The ``exceptional'' orbits at the ``poles'', where one of the two $x_i$'s $= 0$, each thus consist of a single point. We say the orbit lies ``above'' the real number pair that coordinatizes them.

This procedure works in arbitrary dimensions. \begin{defn}
The \emph{toric action} of the $n$-torus $T^n$ on $\CP^n$ is given by the formula 
\[
(\lambda_1, \lambda_2, \ldots, \lambda_n) \cdot (z_0, z_1, \ldots, z_n) = (z_0, \lambda_1 z_1, \ldots, \lambda_n z_n).
\]    
\end{defn} 

Up to global phase, every complex homogeneous coordinate $(z_0, z_1, \ldots, z_n)$ is equivalent to a coordinate of form $(x_0, \lambda_1 x_1, \ldots, \lambda_n x_n)$ with the $x_i$ non-negative real numbers denoting the lengths of the respective complex coordinates $z_i$, the $\lambda_i \in U(1)$, and such that $x_0^2 + x_1^2 + \cdots + x_n^2 = 1$. The proof of the following proposition follows almost exactly as in the case for $\CP^1$.
\begin{theo}
The decomposition of $\CP^n$ into orbits of the toric action of the $n$-torus $T^n$ and points is precisely the decomposition of $\CP^n$ given by declaring states to be equivalent when they behave identically under quantum measurement.    
\end{theo}

Notice that as before this toric geometry decomposition again expresses the toric geometry structure of $\CP^n$ as coordinatized by a set of real convex coordinates for the space of probability distributions over the basis states and a set of periodic coordinates for each of the states in the various toric orbits, the equivalence classes of the states under quantum measurement. These individual periodic coordinates can be thought of as either a real number $\theta \mod 2\pi$ or as the corresponding unit complex number $\lambda = \cos(\theta) + i \sin(\theta)$.  Here we will follow the later convention.

\begin{figure}[H]
    \centering
    \includegraphics[width=0.75\linewidth]{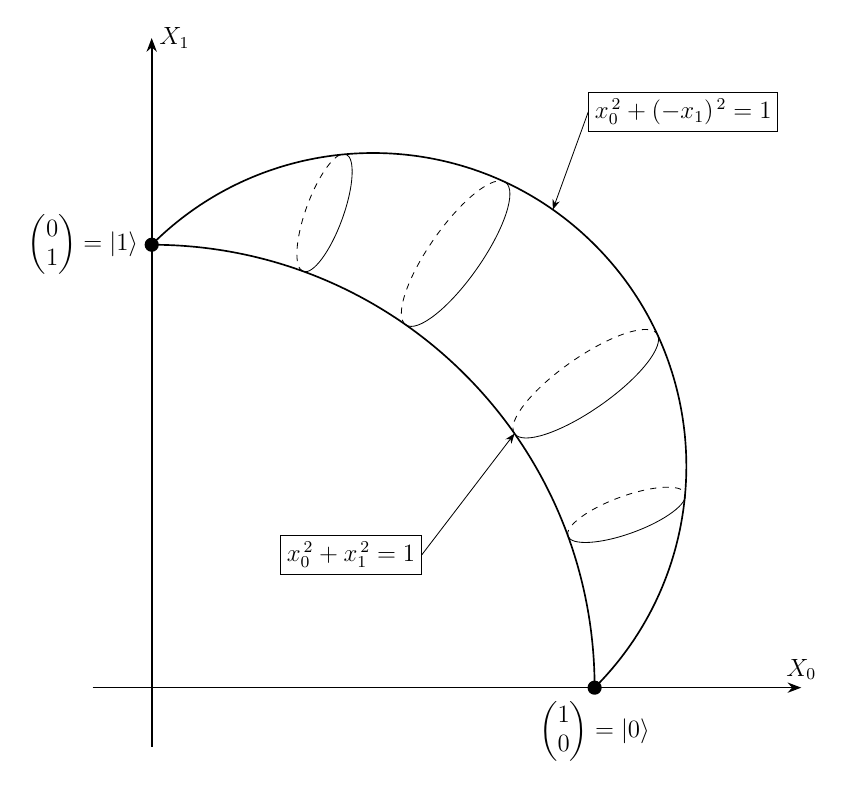}
    \caption{The ``Bloch banana'', showing the circle orbits above the coordinatizing real number pairs.}
    \label{fig:banana}
\end{figure}

It is here that one must face the challenges cartographers have always faced when attempting to express curved objects in flat Euclidean space.  Otherwise, our representations come out unhelpfully curved.  For example, here in Figure~\ref{fig:banana} is a direct expression of $\CP^1$ in toric geometric coordinates, an object several of our colleagues call the ``Bloch banana''.

The ``banana'' has the virtue of better demonstrating the geometry on $\CP^1$ induced by the standard Fubini-Study metric, $\inp{v_0}{v_1} = \bar{v}_0^\Trp v_1$ for affine complex vectors $v_i$, i.e.\ that of a sphere of radius $\frac{1}{2}$, where the angle (i.e.\ the projective distance) between the affine basis elements $(1,0)$ and $(0,1)$ is $\pi/2$ and the length of a great circle is $\pi$.  It also shows the changing geometry of the orbits given by this metric via the decrease in their circumference as they move toward the ``poles''.

That said, the illustration in Figure~\ref{fig:banana} requires some further explanation. The $X_0$ and $X_1$ axes in this figure represent the two real coordinate axes of the two complex coordinates $z_0$ and $z_1$. The two complex coordinates are stylistically represented by the third real axis of the figure. The circles in that figure stylistically represent the set of points $(x_0,\lambda x_1)$ where $\lambda\in U(1)$, the orbit of the state coordinatized by $(x_0,x_1)$ under the action of the 1-torus $T^1\cong U(1)$.

A na\"ive approach to the issues of visualization and curvature is to simply map the points in the Bloch sphere to the probability distribution over the basis elements they represent under quantum measurement.  This takes the states represented in the Bloch sphere to the standard simplex of real convex linear combinations $\Delta^n$ by taking $(x_0, x_1, \ldots, x_n) \mapsto (x_0^2, x_1^2, \ldots, x_n^2)$. While useful for the illustration of certain elementary properties of quantum transformations, this particular map is not linear and thus has the disadvantage of not preserving many of the geometric features of $\CP^n$ as induced by the Fubini-Study metric. In particular, geodesics (i.e.\ straight lines) in $\CP^n$ do not map to straight lines in the standard $n$-simplex $\Delta^n$ under this map.

\begin{figure}[H]
    \includegraphics[width=0.45\linewidth]{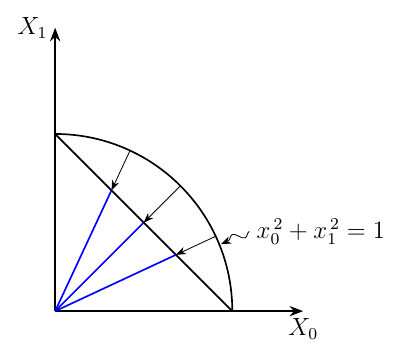}
    \includegraphics[width=0.45\linewidth]{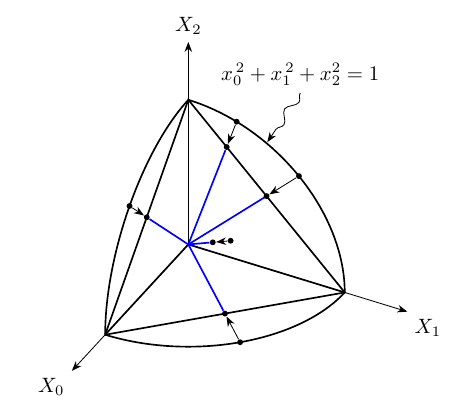}
    \caption{Gnomonic projections in real 2- and 3-dimensional space.}
    \label{fig:gnomonic}
\end{figure}

For more technical geometric analyses, it is useful to employ several of the standard tricks that cartographers have used to express our curved objects in flat Euclidean space.  One such technique is projecting from the center of space to a separate hyperplane in space not through the center (i.e.\ gnomonic projection), the center of the image of the projection being located at the barycenter of the standard simplex. This map is given algebraically by $(x_0,\dots,x_n)$ mapping to $\dfrac{1}{\sum_{k=0}^n x_k}(x_0,\dots,x_n)$. These are the so-called ``gnomonic coordinates'' on the standard simplex as utilized in \cite{bengtsson2002} and illustrated geometrically in dimensions two and three in Figure~\ref{fig:gnomonic}.  Here we gain the geometric property of geodesics mapping to geodesics, i.e.\ straight lines in our model. 

We orient our spaces, and hence our simplices, via the right-hand rule. For the 2-simplex, this gives a counterclockwise orientation of the vertices $\ket{0}$, $\ket{1}$, and $\ket{2}$.

\begin{figure}[H]
    \centering
    \includegraphics[width=0.5\linewidth]{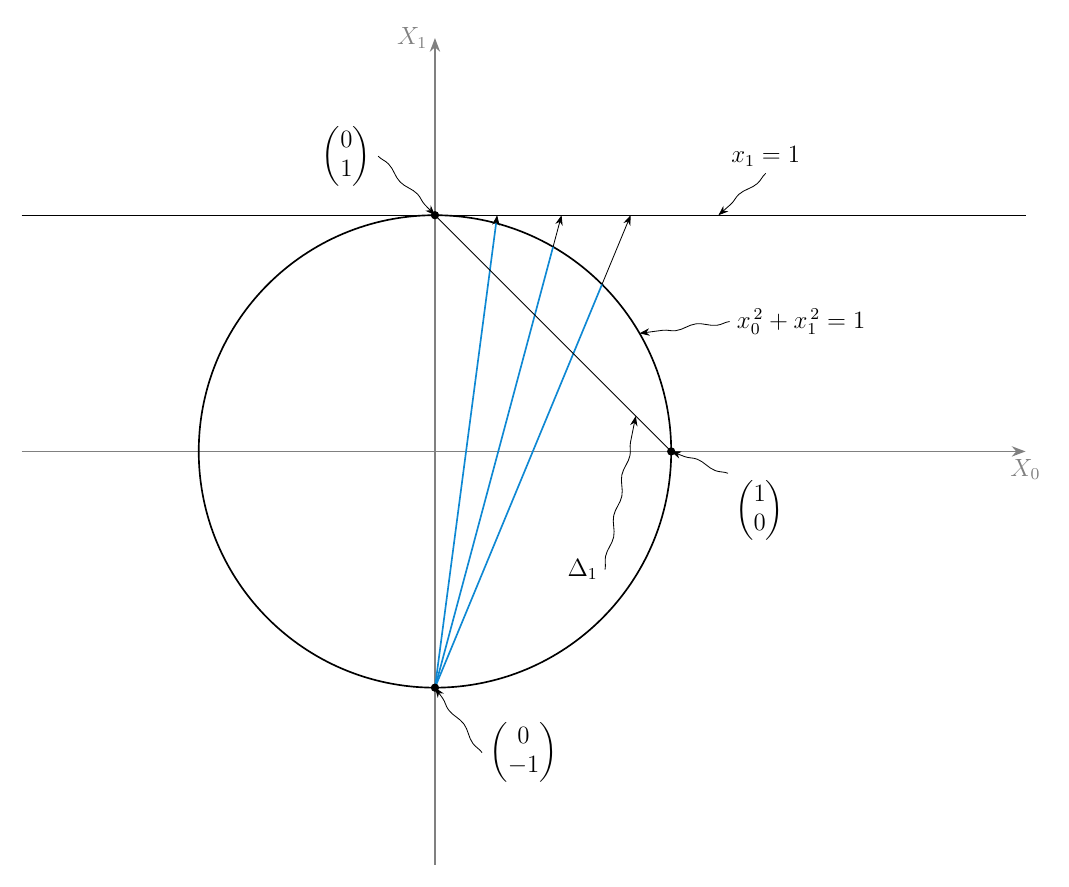}
    \caption{Stereographic projection in real 2-dimensional space, centered at the South pole of the sphere to the tangent line at North pole (top) and to the $X_0$-axis (bottom).}
    \label{fig:stereographic}
\end{figure}

Another useful technique is that of stereographic projection from a point on the sphere to the tangent line at the antipodal point or a diameter parallel to this tangent line. Stereographic projection also maps geodesics to geodesics and can be combined with gnomonic projection. This can also be used to map the non-negative hyperoctant of a sphere to a simplex.  When compared to standalone gnomonic projection, this transformation has the additional geometric property of preserving angles.  For the situation in real 2--dimensional space, see Figure~\ref{fig:stereographic}. The formulae are standard exercises in Euclidean geometry.

\begin{figure}[H]
    \centering
    \includegraphics[width=0.5\linewidth]{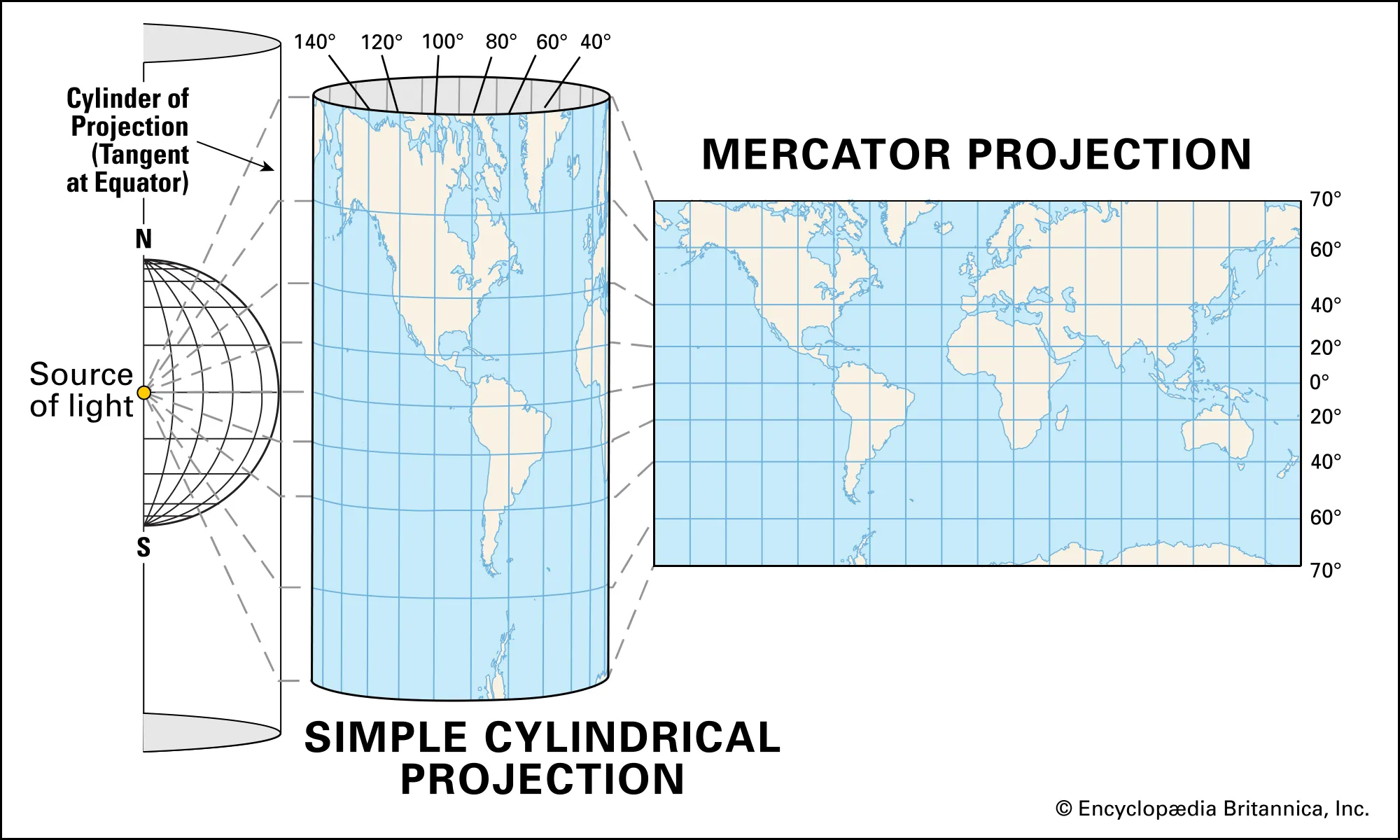}
    \caption{Mercator projection \cite{britannica}.}
    \label{fig:mercator}
\end{figure}

A third technique is the ``cutting open'' of curved objects, thus expressing them as identification spaces, as found in the familiar Mercator projection of the Earth's surface, where the final flat map is obtained by first gnomonically projecting the Earth's surface onto a cylinder tangent to the equator, which is then ``cut open'' at the 180th meridian. The Mercator projection has formulas given as, 
\begin{align*}
    x &= R(\lambda -  \lambda_0)\\ 
     y &= 
        \begin{cases}
            \infty, & \text{if } \phi = \pi / 2 \\
            - \infty, & \text{if } \phi = -\pi / 2\\
            R\cdot\ln\lrs{\tan\lrp{\frac{\pi}{4} +\frac{\phi}{2}}}, & \text{otherwise}
        \end{cases}
\end{align*}  
where R is the radius of the sphere, $\lambda$ is the longitude, and $\phi$ is the latitude \cite{snyder1989}. See Figure~\ref{fig:mercator}. This technique is particularly useful for visualizing geometric tori, which do not embed isometrically in low-dimensional Euclidean space, as illustrated for the 2-torus in Figure~\ref{fig:torus}.

\begin{figure}[H]
    \centering
    \includegraphics[width=0.75\linewidth]{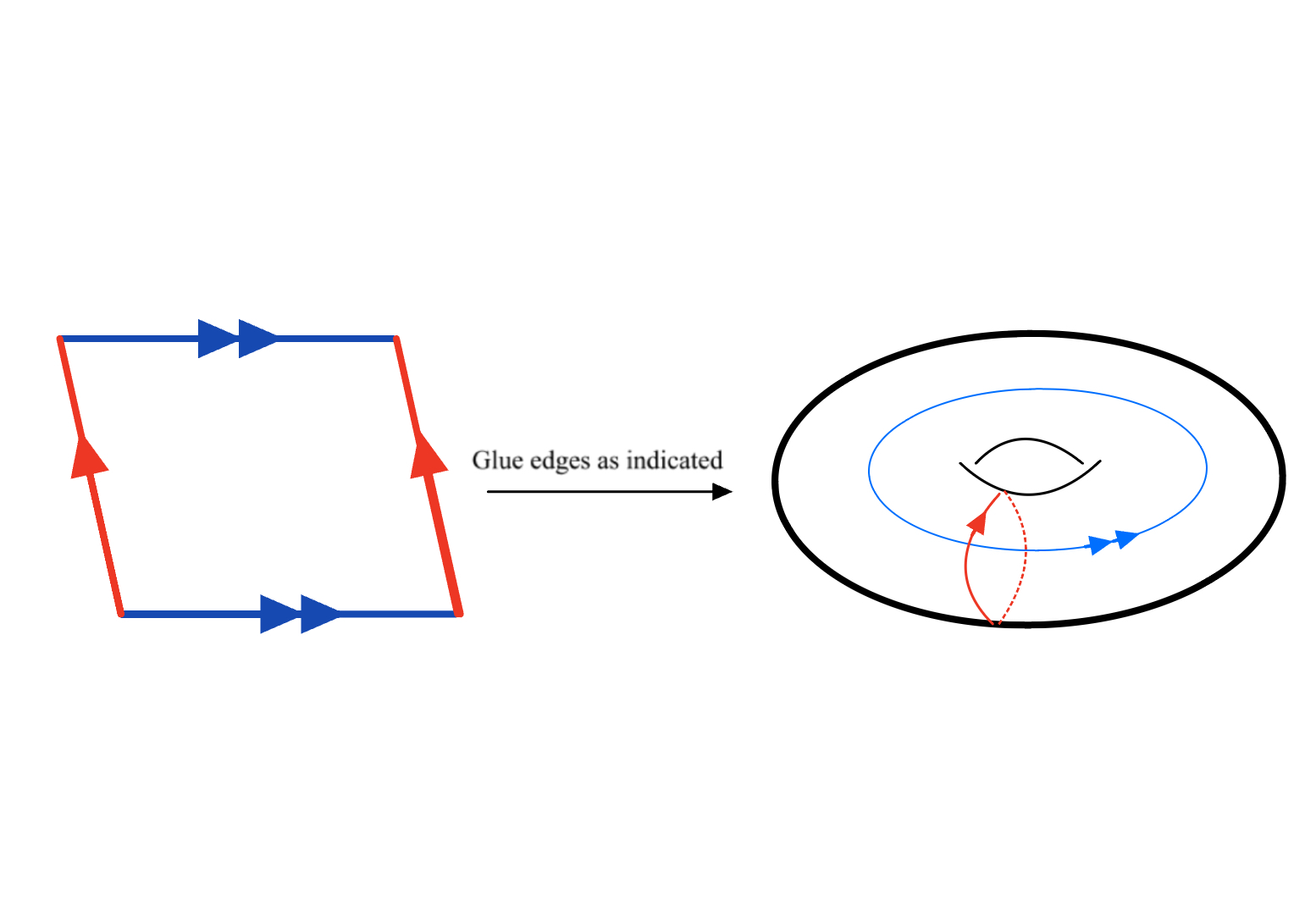}
    \caption{A geometric 2-torus expressed as an identification space with domain a parallelogram, with an angle between longitude and meridian represented at the corners of the parallelogram. In higher dimensions, this parallelogram is replaced by parallelepipeds and their analogues.}
    \label{fig:identification}
\end{figure}

For example, a geometric 2-torus is expressed isometrically as the identification space of a parallelogram, with the lengths of the meridians and longitudes represented by the lengths of the sides in each parallel class of the parallelogram, as depicted in Figure~\ref{fig:identification}.

In particular, under these techniques, while certain distances get distorted, angles may not, and in any event, geodesics in $\CP^n$ map to straight lines in the convex part of our expression of $\CP^n$.  In the periodic part of our expression of $\CP^n$, the ``cutting open'' trick allows for a similar expression of the geometry of the periodic factors.

In this way the edge lengths of the $U(1)$ factors and the angles between them accurately represent the geometric torus orbit lying over the specific convex point that represents the probability distribution over the affine basis elements given by quantum measurement of the quantum state under examination, and an explicit way to analyze the geometric structure of $\CP^n$. 

\begin{figure}[H]
     \centering
     \includegraphics[width=0.75\linewidth]{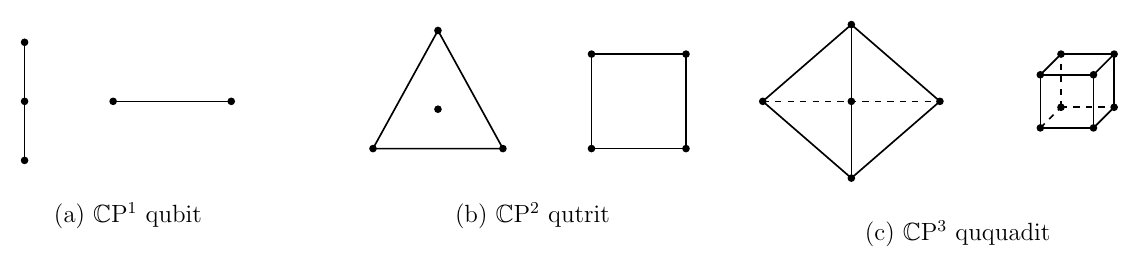}
     \caption{Decompositions of $\CP^1$, $\CP^2$, and $\CP^3$ with ``cut open'' tori (with geometric structure suppressed) whose vertices correspond to an interior point of the simplex.}
     \label{fig:decompositions}
 \end{figure}

As the convex coordinate part of our picture has only $n$ degrees of freedom, this yields an $n + n$ dimensional representation of the $2n$ dimensional $\CP^n$, in a manner similar to the way we considered a cone earlier, see Figure~\ref{fig:decompositions}.

\begin{figure}[H]
    \centering
    \includegraphics[width=0.75\linewidth]{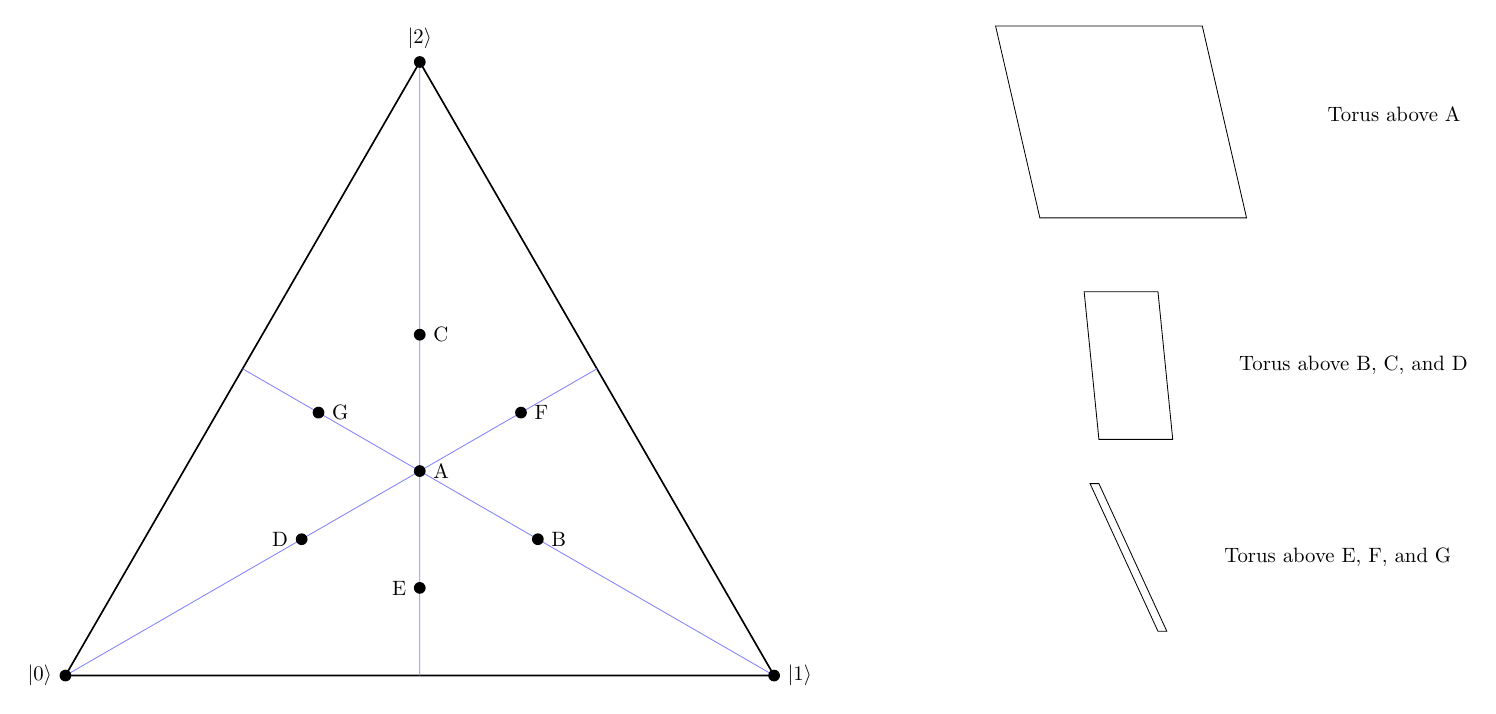}
    \caption{$\CP^2$ with (up to rotation in the plane) the affine geometric structures on the orbit above the various interior points A, B, C, D, E, F, G.}
    \label{fig:cp2orbits}
\end{figure}

For a state with a given probability measure over the basis elements induced by quantum measurement $(x_0,\dots,x_{n})$, the toric geometry orbit of that state has an affine geometry induced by the Fubini-Study metric. This geometric structure changes as the probability distribution over the basis states induced by quantum measurement varies. This structure has a regular affine geometric structure for those states with a uniform distribution over the basis elements given by quantum measurement, for example the barycenter of our simplex. As this probability distribution moves towards the edges, the length of one of the $S^1$ factors decreases, until vanishing at the boundary point, where we now lie in a projective space of one lower dimension. This is illustrated in Figure~\ref{fig:cp2orbits} for $\CP^2$. In higher dimensions, we see the same effect with regards to the length of the edges of the parallelepipeds involved. In $\CP^n$, the general formulae for the parallelepiped side lengths and angles belonging to our identification spaces yielding tori above the point $(x_0,\ldots,x_{n})$ as induced by the Fubini-Study metric are given explicitly by:

\begin{align*}
    L_a = 2\pi \sqrt{x_a} \sqrt{1-x_a}  \quad & \text{and} \quad \theta_{ab}=\arccos\left(\dfrac{\sqrt{x_a x_b}}{\sqrt{1-x_a}\sqrt{1-x_b}}\right)
\end{align*}
with $1\le a\le n$ and $1\le b\le n$.

The reader should note that in our pictures of $\CP^n$ our parallelepipeds and parallelograms degenerate into parallelograms, intervals, and points as we move from interior to face to edge to vertex points on our simplices.  In particular, each edge of our standard simplices in our toric geometry models represents the Bloch sphere formed by the two basis states at the endpoints.  Again, these figures continue to carry the affine geometric structure given by the Fubini-Study metric that depends on the point of the standard simplex these tori project to.

For $\CP^2$, the state space of the quantum trit, this manifests as the affine geometric structure of the parallelogram, which changes from a rhombus above the barycenter to more general parallelograms, in particular with the various coordinate lengths decreasing, as we move toward the edges and vertices of the simplex, see Figure~\ref{fig:cp2orbits}. While not illustrated here, a similar phenomenon occurs in $\CP^3$, the state space of the quantum quadit and joint state space of a pair of qubits, as the affine geometric structure of the corresponding parallelepipeds change from a rhomboid over the barycenter to more general parallelepipeds, as we move toward the faces, edges and vertices of the 3-simplex.

However, for many illustrative purposes used in the analysis of $\CP^n$ and the unitary maps between $\CP^n$ and itself, this part of the information can be safely suppressed, and one can express our tori over interior points of the coordinatizing simplex as identification spaces with domain unit intervals in the case of qubits, squares in the case of qutrits, or cubes in the case of ququadits (joint space of two qubits).

The axes of these identification spaces may be labeled with either complex phase coordinates or angular coordinates, with some transformations more easily expressed using one coordinate system or the other. Multiplication of a complex coordinate by a constant factor corresponds to addition to an angular coordinate by a constant angle, and multiplication or division of powers of complex coordinates corresponds to linear transformation of the angular coordinates within the identification space.

\subsection{Visualizing unitary transformations via toric geometry}

\subsubsection{Visualization}
Fundamental to nearly all binary quantum algorithms is the Hadamard transformation. This is due to the transformation's uniformization properties, that is, this transformation maps each quantum basis state to a uniform superposition of the entire basis. This is what typically allows a parallelism that yields a polynomial speed up over the analogous classical algorithms.  Entanglement can allow a different, exponential speed up. Entanglement, however, requires a circuit, as opposed to just an individual transform. For ternary quantum algorithms, quantum software developers and engineers are faced with the questions of what is the appropriate radix-3 analogue of the uniformizing Hadamard transformation and what are the appropriate ternary quantum circuits to achieve uniformization and entanglement? Similar issues arise for the other commonly used transformations in binary quantum computation.  There is a strong motivation for the determination of transformations directly realizable in hardware. For this, ternary logic can be a more natural setting than binary.  For example, there is a larger information capacity per particle, leading to more compact circuits, there is a greater resilience to certain types of noise, and in the potential for enhanced performance in specific search algorithms. In addition, similar reasoning has been presented for radix-4 \cite{wang2020}.

Our ``map'' of the state space of the qutrit developed in Section~1.2 can be put to good use in the analysis of these questions.  To begin, note that the normalization conventions (multiply a state with non-zero first coordinate by a global phase to make that first coordinate real, or 1) means that in our toric model of the Bloch sphere, rotational transformations which rotate the Bloch sphere around an axis joining the basis states (typically the Z-axis in visualizations of the Bloch sphere) through some angle in $(-\pi, \pi]$ preserve the toric geometric decomposition. The quantum computation community has identified a broad family of such transformations in binary quantum computation, most with rotational angles obtained by dividing $2\pi$ by a suitable power of two.  Examples are the Pauli-Z, S, and T transformations. For radix-3, the increase in the size of our observational basis from two to three says there are now many more such rotational transformations that must be considered, the so-called ``diagonal'' transformations, represented by diagonal unitary matrices, with ``internal'' rotational angles given by dividing $2\pi$ by suitable powers and products of the numbers 2 and 3. 

In all cases, in our visualization of $\CP^n$, any of these ``internal'' rotational transformations of the orbits of the toric action may be visualized simply as a rotation by a fixed angle in one or more of the periodic coordinates of a state, corresponding to a sliding symmetry of the underlying geometric torus or an affine transformation of its fundamental domain.

\begin{figure}[H]
    \centering
    \includegraphics[width=0.75\linewidth]{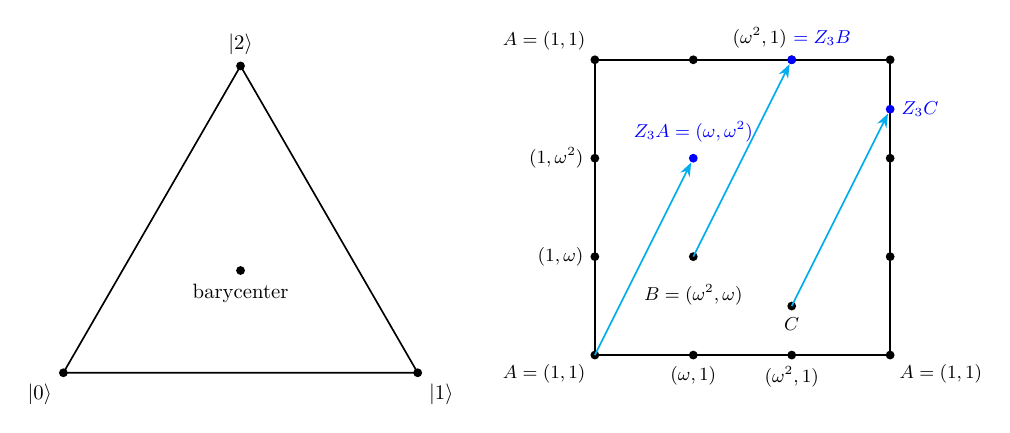}
    \caption{The diagonal map $Z_3=\text{Diag}(1,\omega,\omega^2)$ depicted via its transformation of uniform superpositions of the basis states, which are located in the torus sitting above the barycenter of the convex coordinate simplex, demonstrating an induced affine transformation of the periodic coordinates. The same transformation occurs within all other tori above the other points in the interior of the simplex.}
    \label{fig:rotationmap}
\end{figure}

As an example, in Figure~\ref{fig:rotationmap}, we show the action of the diagonal map $Z_3=\text{Diag}(1, \omega, \omega^2)$ located over the barycenter of the convex coordinate simplex.

If we continue to follow a standard normalization procedure (e.g. normalizing via global phase a non-zero $(1,1)$ entry in a unitary matrix to have value 1 and restricting our coefficients to the cube roots of unity and their additive inverses), we see there are six ``natural'' entries for a diagonal matrix's non-zero coefficients, i.e.\ $\{1, -1, \omega, -\omega, \omega^2, -\omega^2\}$ where $\omega^3 = 1$. There are 36 such possible diagonal matrices given by the six possible entries in each of the $(2,2)$ and $(3,3)$ positions in the matrix. These matrices form a group under composition.  The Cayley graph of this group is of central interest to quantum computation theorists as it can be used to optimize the quantum circuits that employ such transformations.

\begin{figure}[H]
    \centering
    \includegraphics[width=0.75\linewidth]{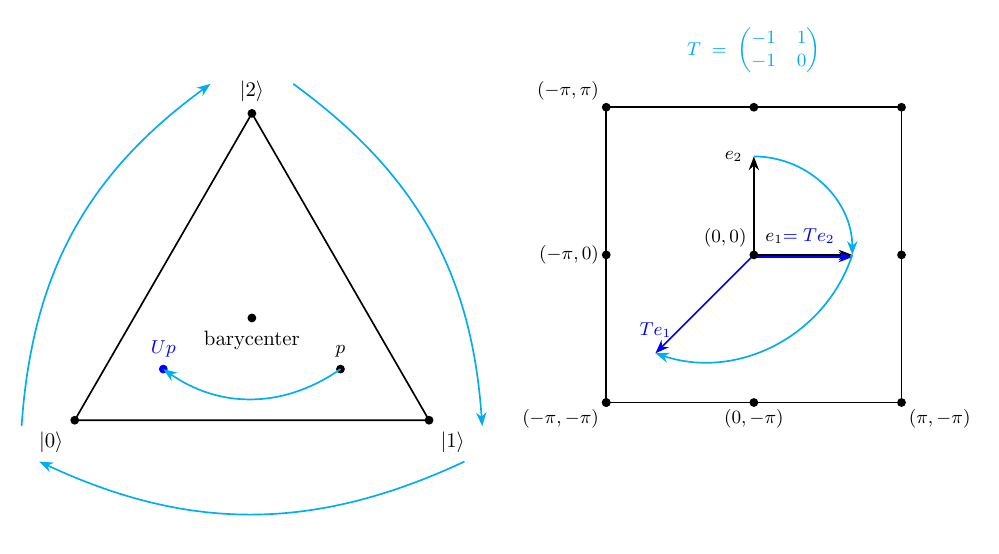}
    \caption{The transformation SHIFT$-1$ depicted via its transformation of the state space $\CP^2$ of the qutrit. Coordinates in the convex coordinate simplex are rotated clockwise by 120 degrees. As our picture requires normalization of the phase of the coefficient of the first basis state to 1, there is an induced transformation $T$ on the periodic coordinates as well. By using angular instead of complex coordinates to locate states within the torus, this induced transformation is realized as a simple linear transformation of the periodic coordinates, visualized here through its action on the basis vectors of the periodic space.}
    \label{fig:shiftminus}
\end{figure}

The ternary permutative transformations also have natural representations in our picture. The permutation $(012)$ is a well-known ternary transformation frequently denoted SHIFT$+1$, similarly the permutation $(021)$ is the well-known ternary transformation frequently denoted SHIFT$+2$, which is occasionally referred to as SHIFT$-1$.
In addition, there are three transformations given by the odd permutations, in particular the transpositions $(01)$, $(02)$, and $(12)$, which interchange the first and second, first and third, and second and third states of the computational basis, respectively. Together with the identity, all 6 of these permutations comprise the well-known symmetric group $S_3$. Within our picture, these permutations are represented by a symmetry -- rotation or reflection, as appropriate -- of the convex coordinate simplex, followed by a linear transformation of the periodic coordinates, which accounts for normalization of the state's first internal phase to 1. Figure~\ref{fig:shiftminus} depicts the transformation SHIFT$-1$ using this method.

This picture holds in general: for an $n+1$-state quantum system, we may apply the toric geometric decomposition to the pure state space $\CP^n$, yielding convex coordinates in the standard simplex $\Delta^n$, with each of the permutative transformations in $S_{n+1}$ as applied to the computational basis elements inducing an isometry of $\Delta^n$ combined with a linear transformation of the periodic coordinates. In particular, for the radix-4 case, our triangle becomes a tetrahedron, our square a cube, and our permutation group has 24 elements. 

\begin{rem} Every product of ``internal'' rotational transformations and permutation transformations of the basis states may be visualized, up to global phase, as an isometry of the convex coordinate simplex followed by an affine transformation of the angular form of the periodic coordinates.\end{rem}

A second fundamental issue concerns the introduction of a natural parallelism into quantum computation, an effect which follows from the ability to place a register of quantum logical units (e.g.\ bits, trits, or quadits) into a uniform superposition of the entire set of basis states for the register.  In binary quantum computation this happy trick is performed by the Hadamard transformation, i.e.\ simultaneously applying the Hadamard transformation to each qubit in the register, having been initialized into the register's lowest energy state.  Of course, the Hadamard transformation is the quantum Fourier transform (QFT) for radix-2 and as a linear transformation has order 2. The other ``uniformizing'' transformations in radix-2 include three other permutative forms of the Hadamard transformation obtained from Hadamard by  applying the permutation NOT in the Domain and range or both As Hadamard has yet to be synthesized as a native gate in any current technology, on occasion these permutative forms of Hadamard can be more natural to employ than Hadamard itself \cite{Ali2025SWAP}. 

Common practice remains that the Hadamard transformation and the resulting transform for qubit registers is used for the ``uniformization'' of the register, even when the transformation needs to be constructed as a circuit from native gates of a particular implementation, as in Qiskit\cite{aleksandrowicz2019qiskit}.

The situation in radix-3 is more complicated.  To begin, the symmetric group $S_3$ consists of two even permutations, the  3-cycles, three transpositions, and the identity. In radix-3 the permutation issue potentially gives 36 permutative forms of a given uniformizing gate. In particular, for the uniformizing gate given by the quantum  Fourier transform in radix-3, also known as the Chrestenson gate, two of these alternate forms were given in \cite{perkowski2007quantum}. These appear in Figure~\ref{fig:chrestenson}. In the engineering literature, these three inequivalent forms of the quantum Fourier transform for radix-3 are called the ``Chrestenson'' transformations \cite{chrestenson1955}, \cite{alrabadi2002}. Each of these permutative variants has order four, and has squares which are permutations representing the three transformations in the symmetric group $S_3$. We note for later use that the square of QFT(3) itself is the permutation (12).

\begin{figure}[H]
    \centering
    \begin{subfigure}{0.3\textwidth}
        \centering
        $\dfrac{1}{\sqrt{3}}
        \begin{bmatrix} 1 & 1 & 1 \\
        1 & \omega & \omega^2 \\ 
        1 & \omega^2 & \omega 
        \end{bmatrix}$
        \caption{QFT(3)}
    \end{subfigure}\hfill
    \begin{subfigure}{0.3\textwidth}
        \centering
        $\dfrac{1}{\sqrt{3}}
        \begin{bmatrix} \omega & 1 & \omega^2 \\
        1 & 1 & 1 \\ 
        \omega^2 & 1 & \omega 
        \end{bmatrix}$
        \caption{Permutation $(012)$}
    \end{subfigure}\hfill
    \begin{subfigure}{0.3\textwidth}
        \centering
        $\dfrac{1}{\sqrt{3}}
        \begin{bmatrix} \omega & \omega^2 & 1 \\
        \omega^2 & \omega & 1 \\
        1 & 1 & 1 
        \end{bmatrix}$
        \caption{Permutation $(021)$}
    \end{subfigure}
    \caption{Matrices for the Chrestenson transformations. (a)~The standard QFT of order~3. (b),~(c)~The QFT obtained by applying the indicated permutation to the basis in both the domain and the range before applying the transformation. The matrix in (a) we will denote by $CH$, the matrix in (b) by $CH_2$, and the matrix in (c) by $CH_3$.}
    \label{fig:chrestenson}
\end{figure}

There are 36 such potential modifications of the quantum Fourier transform for 
radix-3, one for each pair $(\pi_D,\pi_R)\in S_3\times S_3$ of permutations applied to the domain and range bases respectively. Because the QFT(3) matrix equals its transpose, the pairs $(\pi_D,\pi_R)$ and $(\pi_R,\pi_D)$ yield matrices
that are transposes of each other, and hence only 18 distinct matrices arise;
the full enumeration is given in Table~\ref{tab:qft3-forms}.

\begin{table}[h]
    \centering
    \small
    \setlength{\tabcolsep}{4pt}
    \renewcommand{\arraystretch}{1.15}
    \setlength{\arraycolsep}{3pt}
    \begin{tabular}{c|cccccc}
         & $\pi_R=e$ & $\pi_R=(01)$ & $\pi_R=(02)$ & $\pi_R=(12)$ & $\pi_R=(012)$ & $\pi_R=(021)$ \\
        \hline
        $\pi_D=e$ &
            $\begin{bmatrix}1&1&1\\1&\omega&\omega^2\\1&\omega^2&\omega\end{bmatrix}$ &
            $\begin{bmatrix}1&1&1\\\omega&1&\omega^2\\\omega^2&1&\omega\end{bmatrix}$ &
            $\begin{bmatrix}1&1&1\\\omega^2&\omega&1\\\omega&\omega^2&1\end{bmatrix}$ &
            $\begin{bmatrix}1&1&1\\1&\omega^2&\omega\\1&\omega&\omega^2\end{bmatrix}$ &
            $\begin{bmatrix}1&1&1\\\omega^2&1&\omega\\\omega&1&\omega^2\end{bmatrix}$ &
            $\begin{bmatrix}1&1&1\\\omega&\omega^2&1\\\omega^2&\omega&1\end{bmatrix}$ \\[2.3em]
        $\pi_D=(01)$ &
            $\begin{bmatrix}1&\omega&\omega^2\\1&1&1\\1&\omega^2&\omega\end{bmatrix}$ &
            $\begin{bmatrix}\omega&1&\omega^2\\1&1&1\\\omega^2&1&\omega\end{bmatrix}$ &
            $\begin{bmatrix}\omega^2&\omega&1\\1&1&1\\\omega&\omega^2&1\end{bmatrix}$ &
            $\begin{bmatrix}1&\omega^2&\omega\\1&1&1\\1&\omega&\omega^2\end{bmatrix}$ &
            $\begin{bmatrix}\omega^2&1&\omega\\1&1&1\\\omega&1&\omega^2\end{bmatrix}$ &
            $\begin{bmatrix}\omega&\omega^2&1\\1&1&1\\\omega^2&\omega&1\end{bmatrix}$ \\[2.3em]
        $\pi_D=(02)$ &
            $\begin{bmatrix}1&\omega^2&\omega\\1&\omega&\omega^2\\1&1&1\end{bmatrix}$ &
            $\begin{bmatrix}\omega^2&1&\omega\\\omega&1&\omega^2\\1&1&1\end{bmatrix}$ &
            $\begin{bmatrix}\omega&\omega^2&1\\\omega^2&\omega&1\\1&1&1\end{bmatrix}$ &
            $\begin{bmatrix}1&\omega&\omega^2\\1&\omega^2&\omega\\1&1&1\end{bmatrix}$ &
            $\begin{bmatrix}\omega&1&\omega^2\\\omega^2&1&\omega\\1&1&1\end{bmatrix}$ &
            $\begin{bmatrix}\omega^2&\omega&1\\\omega&\omega^2&1\\1&1&1\end{bmatrix}$ \\[2.3em]
        $\pi_D=(12)$ &
            $\begin{bmatrix}1&1&1\\1&\omega^2&\omega\\1&\omega&\omega^2\end{bmatrix}$ &
            $\begin{bmatrix}1&1&1\\\omega^2&1&\omega\\\omega&1&\omega^2\end{bmatrix}$ &
            $\begin{bmatrix}1&1&1\\\omega&\omega^2&1\\\omega^2&\omega&1\end{bmatrix}$ &
            $\begin{bmatrix}1&1&1\\1&\omega&\omega^2\\1&\omega^2&\omega\end{bmatrix}$ &
            $\begin{bmatrix}1&1&1\\\omega&1&\omega^2\\\omega^2&1&\omega\end{bmatrix}$ &
            $\begin{bmatrix}1&1&1\\\omega^2&\omega&1\\\omega&\omega^2&1\end{bmatrix}$ \\[2.3em]
        $\pi_D=(012)$ &
            $\begin{bmatrix}1&\omega^2&\omega\\1&1&1\\1&\omega&\omega^2\end{bmatrix}$ &
            $\begin{bmatrix}\omega^2&1&\omega\\1&1&1\\\omega&1&\omega^2\end{bmatrix}$ &
            $\begin{bmatrix}\omega&\omega^2&1\\1&1&1\\\omega^2&\omega&1\end{bmatrix}$ &
            $\begin{bmatrix}1&\omega&\omega^2\\1&1&1\\1&\omega^2&\omega\end{bmatrix}$ &
            $\begin{bmatrix}\omega&1&\omega^2\\1&1&1\\\omega^2&1&\omega\end{bmatrix}$ &
            $\begin{bmatrix}\omega^2&\omega&1\\1&1&1\\\omega&\omega^2&1\end{bmatrix}$ \\[2.3em]
        $\pi_D=(021)$ &
            $\begin{bmatrix}1&\omega&\omega^2\\1&\omega^2&\omega\\1&1&1\end{bmatrix}$ &
            $\begin{bmatrix}\omega&1&\omega^2\\\omega^2&1&\omega\\1&1&1\end{bmatrix}$ &
            $\begin{bmatrix}\omega^2&\omega&1\\\omega&\omega^2&1\\1&1&1\end{bmatrix}$ &
            $\begin{bmatrix}1&\omega^2&\omega\\1&\omega&\omega^2\\1&1&1\end{bmatrix}$ &
            $\begin{bmatrix}\omega^2&1&\omega\\\omega&1&\omega^2\\1&1&1\end{bmatrix}$ &
            $\begin{bmatrix}\omega&\omega^2&1\\\omega^2&\omega&1\\1&1&1\end{bmatrix}$ \\
    \end{tabular}
    \caption{The 18 forms of the radix-3 QFT, indexed by the domain
    permutation $\pi_D$ (rows) and range permutation $\pi_R$ (columns).
    The global factor $\tfrac{1}{\sqrt{3}}$ has been omitted from every
    entry. The diagonal cells $(e,e)$, $((012),(012))$, and $((021),(021))$
    reproduce the three matrices of Figure~\ref{fig:chrestenson}.}
    \label{tab:qft3-forms}
    
\end{table}

\begin{figure}
    \centering
    \includegraphics[width=0.75\linewidth]{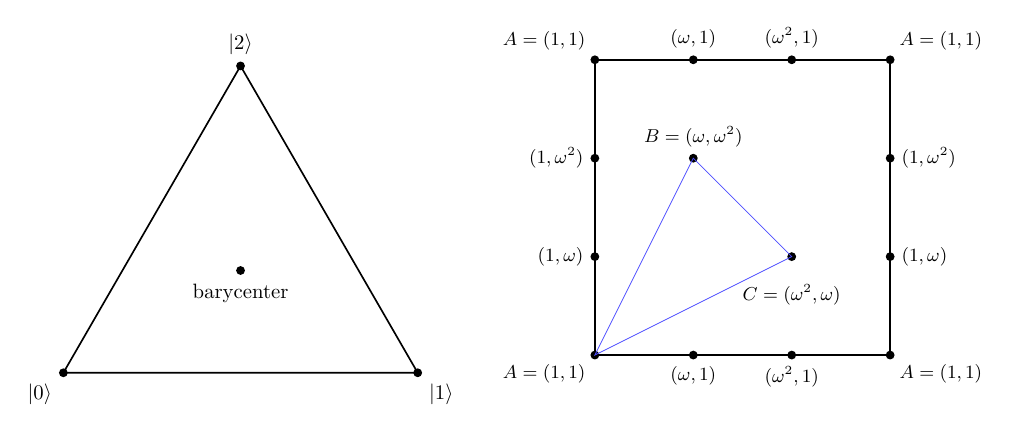}
    \caption{Images of the convex coordinates under the three radix-3 quantum Fourier transforms (Chrestenson transformations) in the 2-torus ``above'' the barycenter, i.e.\ the uniform distribution over the basis states.}
    \label{fig:chrestenson_images}
\end{figure}

In our toric geometric visualization  of the state space of the qutrit, all of the Chrestenson transformation variants map the set of basis states (and hence the entire simplex) into the torus ``above'' the barycenter of our 2-simplex of probability distributions over the basis states as shown in Figure~\ref{fig:chrestenson_images}.  Specifically, for QFT(3) the state represented by $\ket{0} + \ket{1} + \ket{2}$ appears as the point A, the state represented by $\ket{0} + \omega\ket{1} + \omega^2\ket{2}$ as B, and the state represented by $\ket{0} + \omega^2\ket{1} + \omega\ket{2}$ as the point C.  For example, for the two other Chrestenson transformations  of Figure~\ref{fig:chrestenson} this triangle is rotated clockwise through an angle of $2\pi/3$ with each application of the permutation $(012)$ to the bases of both the domain and range of the transformation.

For the other Chrestenson transformation variants, this triangle is mapped via a symmetry corresponding to the maps of the vertices as given by the specific permutations applied to the domain and range of the transformation.
\clearpage

\begin{figure}[H]
    \centering
    \includegraphics[width=0.75\linewidth]{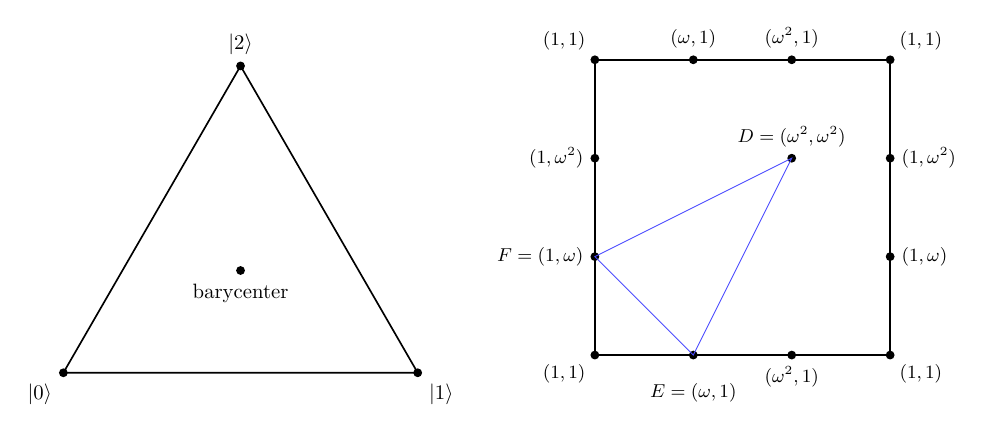}
    \caption{Images of the orbit space under $SB$ in the 2-torus ``above'' the barycenter, i.e.\ the uniform distribution over the basis states.}
    \label{fig:nonqft1}
\end{figure}

\begin{figure}[H]
    \centering
    \includegraphics[width=0.75\linewidth]{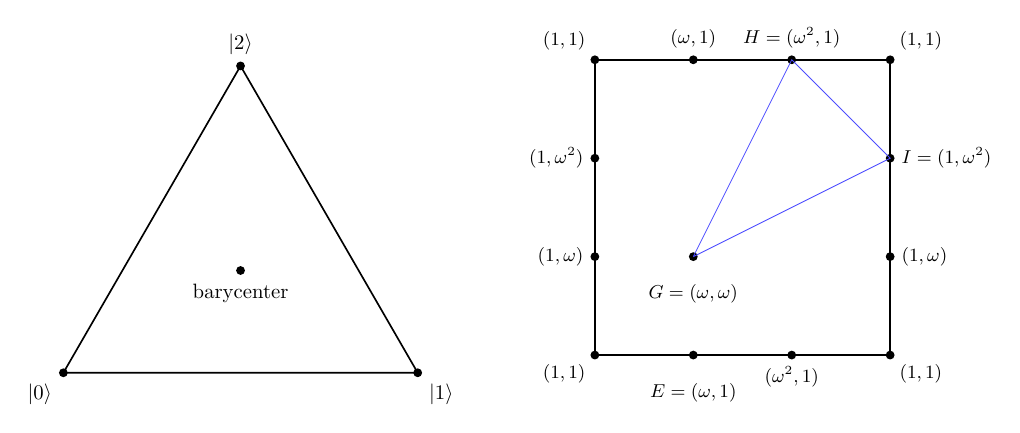}
    \caption{Images of the orbit space under $SB^\dagger$ in the 2-torus ``above'' the barycenter, i.e. the uniform distribution over the basis states.}
    \label{fig:nonqft2}
\end{figure}

In radix-3 another issue arises not present in radix-2. In radix-3 there are other unitary transformations with their corresponding permutative forms, that ``uniformize'' superpositions in a manner similar to the Chrestenson transformations. As an example, we give in Figure~\ref{fig:nonqft_matrices} such a unitary transformation, which we call $SB$ and its inverse $SB^\dagger$, along with two of their permutative variants, as described for the Chrestenson transformation above. See Figures~\ref{fig:nonqft_matrices}, \ref{fig:nonqft1}, and \ref{fig:nonqft2}. To give the reader a sense of these permutative  variants, in those figures we apply the permutations (012) and (021) to both the domain and range bases as in \cite{perkowski2007quantum}. 

\begin{figure}[H]
    \centering
    \[
        SB=\frac{1}{\sqrt{3}}
        \begin{bmatrix} 1 & 1 & \omega \\ 1 & \omega & 1 \\ \omega & 1 & 1 \end{bmatrix},\;
        \frac{1}{\sqrt{3}}
        \begin{bmatrix} 1 & \omega & 1 \\ \omega & 1 & 1 \\ 1 & 1 & \omega \end{bmatrix},\;
        \frac{1}{\sqrt{3}}
        \begin{bmatrix} \omega & 1 & 1 \\ 1 & 1 & \omega \\ 1 & \omega & 1 \end{bmatrix}
    \]
    and
    \[
        SB^\dagger = \frac{1}{\sqrt{3}}
        \begin{bmatrix} 1 & 1 & \omega^2 \\ 1 & \omega^2 & 1 \\ \omega^2 & 1 & 1 \end{bmatrix},\;
        \frac{1}{\sqrt{3}}
        \begin{bmatrix} 1 & \omega^2 & 1 \\ \omega^2 & 1 & 1 \\ 1 & 1 & \omega^2 \end{bmatrix},\;
        \frac{1}{\sqrt{3}}
        \begin{bmatrix} \omega^2 & 1 & 1 \\ 1 & 1 & \omega^2 \\ 1 & \omega^2 & 1 \end{bmatrix}
    \]
    \caption{Two non-QFT uniformizing unitary transformations $SB$, $SB^\dagger$ and a pair of permutative variants of each for radix-3.}
    \label{fig:nonqft_matrices}
\end{figure}
\clearpage

These uniformizing transformations and their permutative variants are very different than the Chrestenson transformations in that the Chrestenson transformation has order $4$ and 18 distinct permutative variants, while $SB$ has order $12$ and 12 distinct permutative variants, as shown in Table~\ref{tab:non-qft-uniformizing-forms}.

\begin{table}[h] 
    \centering
    \small
   \setlength{\tabcolsep}{4pt}
    \renewcommand{\arraystretch}{1.15}
    \setlength{\arraycolsep}{3pt}
    \begin{tabular}{c|cccccc}
         & $\pi_R=e$ & $\pi_R=(01)$ & $\pi_R=(02)$ & $\pi_R=(12)$ & $\pi_R=(012)$ & $\pi_R=(021)$ \\
        \hline
        $\pi_D=e$ &
            $\begin{bmatrix}1&1&\omega\\1&\omega&1\\\omega&1&1\end{bmatrix}$ &
            $\begin{bmatrix}1&1&\omega\\\omega&1&1\\1&\omega&1\end{bmatrix}$ &
            $\begin{bmatrix}\omega&1&1\\1&\omega&1\\1&1&\omega\end{bmatrix}$ &
            $\begin{bmatrix}1&\omega&1\\1&1&\omega\\\omega&1&1\end{bmatrix}$ &
            $\begin{bmatrix}\omega&1&1\\1&1&\omega\\1&\omega&1\end{bmatrix}$ &
            $\begin{bmatrix}1&\omega&1\\\omega&1&1\\1&1&\omega\end{bmatrix}$ \\[2.3em]
        $\pi_D=(01)$ &
            $\begin{bmatrix}1&\omega&1\\1&1&\omega\\\omega&1&1\end{bmatrix}$ &
            $\begin{bmatrix}\omega&1&1\\1&1&\omega\\1&\omega&1\end{bmatrix}$ &
            $\begin{bmatrix}1&\omega&1\\\omega&1&1\\1&1&\omega\end{bmatrix}$ &
            $\begin{bmatrix}1&1&\omega\\1&\omega&1\\\omega&1&1\end{bmatrix}$ &
            $\begin{bmatrix}1&1&\omega\\\omega&1&1\\1&\omega&1\end{bmatrix}$ &
            $\begin{bmatrix}\omega&1&1\\1&\omega&1\\1&1&\omega\end{bmatrix}$ \\[2.3em]
        $\pi_D=(02)$ &
            $\begin{bmatrix}\omega&1&1\\1&\omega&1\\1&1&\omega\end{bmatrix}$ &
            $\begin{bmatrix}1&\omega&1\\\omega&1&1\\1&1&\omega\end{bmatrix}$ &
            $\begin{bmatrix}1&1&\omega\\1&\omega&1\\\omega&1&1\end{bmatrix}$ &
            $\begin{bmatrix}\omega&1&1\\1&1&\omega\\1&\omega&1\end{bmatrix}$ &
            $\begin{bmatrix}1&\omega&1\\1&1&\omega\\\omega&1&1\end{bmatrix}$ &
            $\begin{bmatrix}1&1&\omega\\\omega&1&1\\1&\omega&1\end{bmatrix}$ \\[2.3em]
        $\pi_D=(12)$ &
            $\begin{bmatrix}1&1&\omega\\\omega&1&1\\1&\omega&1\end{bmatrix}$ &
            $\begin{bmatrix}1&1&\omega\\1&\omega&1\\\omega&1&1\end{bmatrix}$ &
            $\begin{bmatrix}\omega&1&1\\1&1&\omega\\1&\omega&1\end{bmatrix}$ &
            $\begin{bmatrix}1&\omega&1\\\omega&1&1\\1&1&\omega\end{bmatrix}$ &
            $\begin{bmatrix}\omega&1&1\\1&\omega&1\\1&1&\omega\end{bmatrix}$ &
            $\begin{bmatrix}1&\omega&1\\1&1&\omega\\\omega&1&1\end{bmatrix}$ \\[2.3em]
        $\pi_D=(012)$ &
            $\begin{bmatrix}\omega&1&1\\1&1&\omega\\1&\omega&1\end{bmatrix}$ &
            $\begin{bmatrix}1&\omega&1\\1&1&\omega\\\omega&1&1\end{bmatrix}$ &
            $\begin{bmatrix}1&1&\omega\\\omega&1&1\\1&\omega&1\end{bmatrix}$ &
            $\begin{bmatrix}\omega&1&1\\1&\omega&1\\1&1&\omega\end{bmatrix}$ &
            $\begin{bmatrix}1&\omega&1\\\omega&1&1\\1&1&\omega\end{bmatrix}$ &
            $\begin{bmatrix}1&1&\omega\\1&\omega&1\\\omega&1&1\end{bmatrix}$ \\[2.3em]
        $\pi_D=(021)$ &
            $\begin{bmatrix}1&\omega&1\\\omega&1&1\\1&1&\omega\end{bmatrix}$ &
            $\begin{bmatrix}\omega&1&1\\1&\omega&1\\1&1&\omega\end{bmatrix}$ &
            $\begin{bmatrix}1&\omega&1\\1&1&\omega\\\omega&1&1\end{bmatrix}$ &
            $\begin{bmatrix}1&1&\omega\\\omega&1&1\\1&\omega&1\end{bmatrix}$ &
            $\begin{bmatrix}1&1&\omega\\1&\omega&1\\\omega&1&1\end{bmatrix}$ &
            $\begin{bmatrix}\omega&1&1\\1&1&\omega\\1&\omega&1\end{bmatrix}$ \\
    \end{tabular}
    \caption{The 36 potential permutative forms of the uniformizing unitary transformation $SB$, indexed by domain permutation $\pi_D$ (rows) and range permutation $\pi_R$ (columns). The global factor $\frac{1}{\sqrt{3}}$ is suppressed. Only 6 distinct transformations occur (each appearing 6 times). Replacing $\omega$ by $\omega^2$ in the matrices above yields 6 distinct permutative variants for $SB^\dagger$ and hence a total of 12 permutative variants for $SB$.}
    \label{tab:non-qft-uniformizing-forms}
\end{table}

In all our pictures of uniformizing transformations, note that the interiors of the triangles are the images of the probability distributions over the basis states. Also note the contrast with radix-2 where the single Hadamard transform, possibly followed by Pauli-Z rotations, is universally employed to uniformize the state of a register of qubits. In radix-3 we find there is a choice to be made between many uniformization transformations or combinations thereof.

One issue here for the engineers involved in the technological design of gates is: Which of these many transformations (or combinations thereof) can be cost effectively realized in hardware?

Similar issues for the engineers arise when considering the ternary analogues of the singly and multiply controlled NOT transformations (i.e.\ the CNOT and Toffoli transformations) that are regularly used in binary quantum algorithms.  For example, there are now more choices of control states, and the transposition of basis states given by NOT must be replaced by one of the six elements of the symmetric group $S_3$. As with uniformization, rather than a single ``useful'' Toffoli transformation with four possible control states, there are many more transformations that perform a Toffoli-like function in ternary, with nine control states, and for each specific purpose, choices must be made. In ternary quantum circuits, the target of a control transformation takes on a heightened importance. This is because many more transformations occur on the target qutrit line as the single qutrit transformation under control frequently needs to be conjugated by other single qutrit transformations. 

The ternary circuits obtained via the visualization of quantum trits introduced here realize various ternary logics, such as the Galois field 3 logic, Reed-Muller 3 (also called Ternary Reed-Muller) \cite{green1989ternary,lee1999generation}, and Post and {\L}ukasiewicz logic \cite{muzio1986}.  

\subsubsection{Transformations for Ternary Quantum Logic}\label{sec:ternaryquantumlogictransforms}

A collection of six ternary single-qutrit and 15 two-qutrit permutative transformations is postulated in several papers in which the universality of the collection for ternary quantum logic is also established \cite{muthukrishnan2000multivalued,deibuk2015optimized}. From these 21 transformations certain ternary circuits are mathematically derived \cite{khan2007quantum}, for example arithmetic circuits such as adders \cite{khan2007quantumJSA,bocharov2015improved}. However, the authors referenced attempted to find a minimized set of basic transformations from which these 21 described transformations can be implemented, either via simulation or hardware, with varying results. We call this \emph{the permutative circuit notation}. Here we propose several sets of transformations, considered as basic gates, minimal with respect to universality for permutative ternary quantum circuits. Additionally, we propose larger, including maximal, sets of basic transformations from which more complex gates can be synthesized. These are described below. Exploiting the visualization given via the toric geometric structure on the complex projective spaces described herein, we indicate how to build practically realizable permutative gates using basic rotations. Subsequently the 21 transformations described above, proved to be universal in \cite{muthukrishnan2000multivalued}, can then be practically realized via quantum circuits synthesized from these basic transformations. Existing quantum programming systems, e.g. IBM Qiskit \cite{aleksandrowicz2019qiskit,javadi2024quantum}, include a set of basic binary quantum transformations. These sets are universal for binary quantum computation and so allow the synthesis of arbitrary binary quantum circuits, including permutative and uniformizing circuits. These latter circuits, for example the Hadamard transform in binary quantum computation, are fundamental to the polynomial speedup found in many quantum algorithms.

\section{Engineering Applications}

The principal contributions of the next few sections are the development of sets of quantum gates universal for permutative ternary quantum circuits. Several of these gates are realized in hardware in various specific implementation technologies, such as superconducting or optical~\cite{goss2022high,thornton2018radix4chrestenson}. Other gates in our sets are realized in circuits using those realized in hardware gates exclusively. Each set serves as a universal set for permutative ternary quantum circuits. We begin by proposing a minimal set of gates that generate all the ternary permutative transformations. Subsequently, we propose a non-minimal set of gates that is universal for permutative ternary quantum circuits that allows for more efficient circuit design.

\subsection{A minimal universal gate set for permutative ternary quantum circuits realizable in current technologies}

In ternary logic, analogues to the Pauli rotations of binary logic are expressed via the so-called ``diagonal transformations''. These transformations, when combined with uniformizing transformations such as the Chrestenson and other transforms described above, form a complete basis for permutative ternary quantum circuits, similar to the binary case as described in the previous section. One such diagonal gate generalizing the Pauli-Z transform in binary logic to ternary is given by $\text{Diag}(1,\omega,\omega^2)$, which we indicate as the $Z_3$ gate and display explicitly in Figure~\ref{fig:20_gates}. 

\begin{figure}[H]
  \centering
   \includegraphics[width=0.90\linewidth]{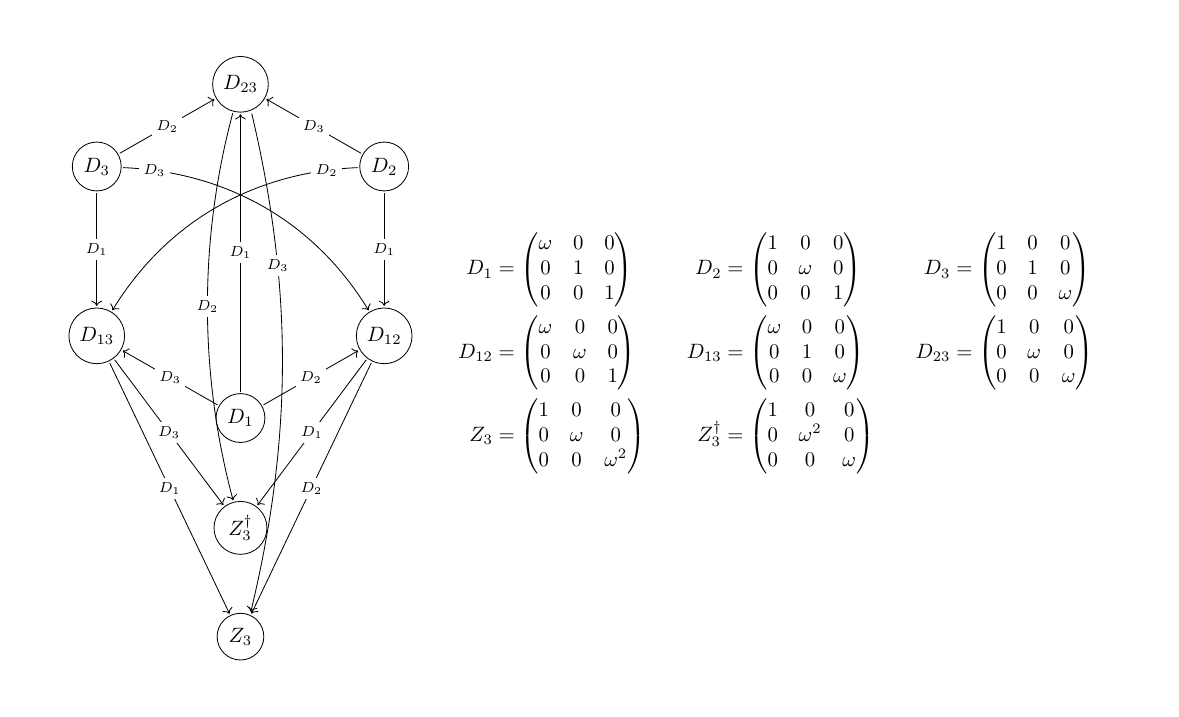}   
   \caption{A subgraph of the Cayley graph of the group of diagonal gates.}
   \label{fig:cayleygraph1}
\end{figure}

A useful mathematical tool for the determination of the least expensive and most reliable circuits for a given unitary transformation is the Cayley graph or subgraphs thereof, of the group of diagonal transformations. For those diagonal transformations with coefficients taken from the set of principal cube roots of unity, this subgraph is given in Figure~\ref{fig:cayleygraph1}. In Figure~\ref{fig:cayleygraph1}, there is a global phase (coefficient) difference between the diagonal transformations of all the diagonal matrices, rather than the direct result of matrix multiplication, e.g., $D_1 \cdot D_1 = \omega D_{23}$.

For later use, we introduce into this subgraph permutative variants of the Chrestenson gates as illustrated in Figure~\ref{fig:chrestenson}. We will use the relations indicated in this subgraph in our development of minimal universal gate sets for permutative ternary quantum circuits.

\begin{figure}[H]
   \centering
   \includegraphics[width=0.8\linewidth]{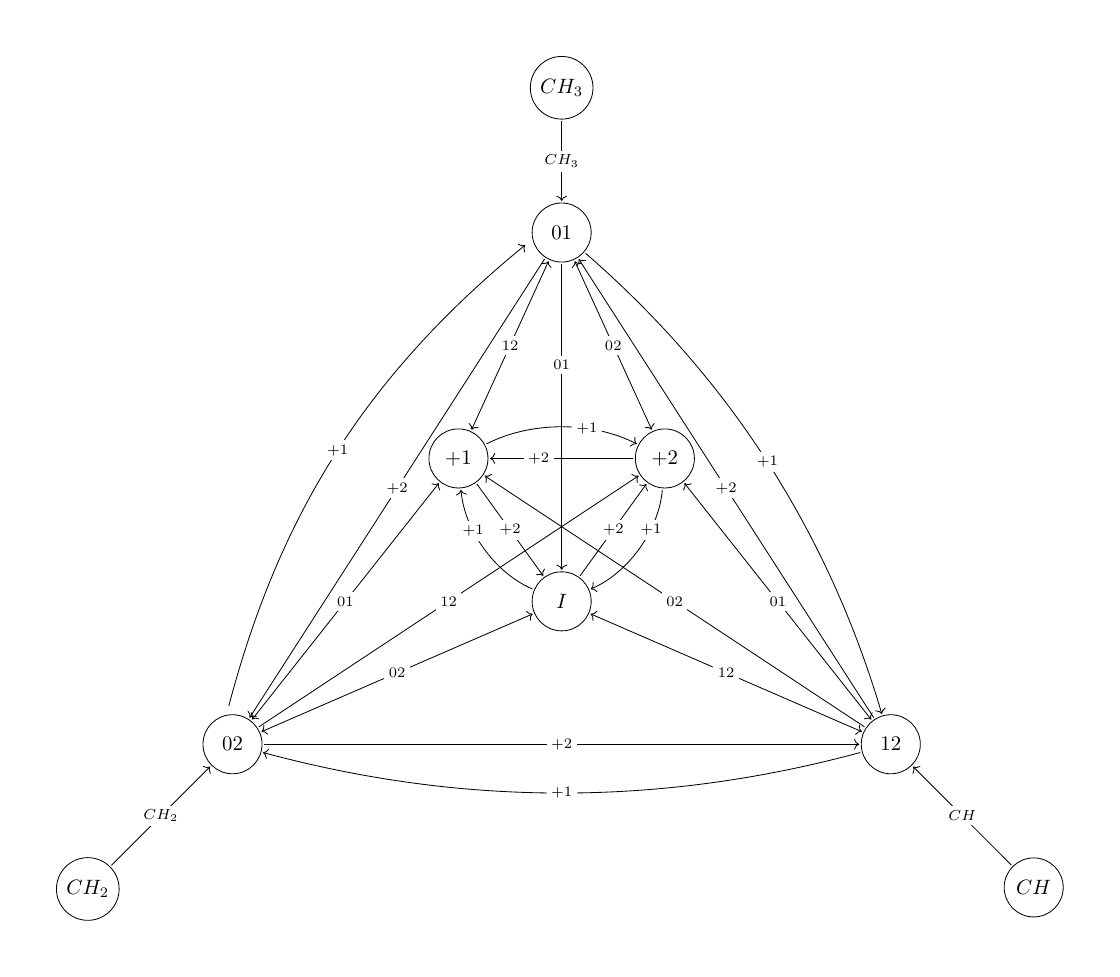}   
   \caption{A subgraph of the Cayley graph of the group generated by permutative and Chrestenson gates.}
   \label{fig:cayleygraph2}
\end{figure}

For this development of universal gate sets for permutative ternary quantum circuits, we begin by considering the following set of unitary transformations given in matrix form by,

\begin{figure}[H]
\centering
\[
CH= \frac{1}{\sqrt{3}}\begin{bmatrix} 1 & 1 & 1 \\ 1 & \omega & \omega^2 \\ 1 & \omega^2 & \omega \end{bmatrix} , \quad
Z_3 =\begin{bmatrix}1 & 0 & 0 \\ 0 & \omega & 0 \\ 0 & 0 & \omega^2\end{bmatrix}.
\]
\caption{}
\label{fig:20_gates}
\end{figure}

The first transformation in our collection is the well-known ternary quantum Fourier transform, also known in the literature with various names, including the Chrestenson , ternary Hadamard, and Chrestenson-Vilenkin gate. The second transformation in our collection is a commonly used ternary analog to the Pauli-Z rotation in binary quantum computation. The $Z_3$ gate is formed by placing the cube roots of unity along the diagonal in a manner similar to the matrix of the Pauli-Z rotation formed by placing the square roots of unity along the diagonal.

\begin{theo}
The gate set, $\lrb{CH, Z_3}$ forms a universal generating set for permutative ternary quantum circuits.
\end{theo}
\begin{proof}  
To begin, note that the permutative transform $(12) = CH \cdot CH$ :
\begin{equation*}
    \begin{split}
        (12) 
        & = \dfrac{1}{\sqrt{3}}\begin{bmatrix} 1 & 1 & 1 \\ 1 & \omega & \omega^2 \\ 1 & \omega^2 & \omega \end{bmatrix} \dfrac{1}{\sqrt{3}} \begin{bmatrix} 1 & 1 & 1 \\ 1 & \omega & \omega^2 \\ 1 & \omega^2 & \omega \end{bmatrix}
        = \begin{bmatrix}
            1&0&0\\
            0&0&1\\
            0&1&0
        \end{bmatrix}
    \end{split}  
\end{equation*}

Similarly, we get $(02)  = CH \cdot Z_3 \cdot CH$:
 
\begin{equation*}
    \begin{split}
        (02)
        & = \frac{1}{\sqrt{3}}\begin{bmatrix} 1 & 1 & 1 \\ 1 & \omega & \omega^2 \\ 1 & \omega^2 & \omega \end{bmatrix} \begin{bmatrix}1 & 0 & 0 \\ 0 & \omega & 0 \\ 0 & 0 & \omega^2\end{bmatrix}\frac{1}{\sqrt{3}}\begin{bmatrix} 1 & 1 & 1 \\ 1 & \omega & \omega^2 \\ 1 & \omega^2 & \omega \end{bmatrix} = \begin{bmatrix}
            0 & 0 & 1 \\ 0 & 1 & 0 \\ 1 & 0 & 0
        \end{bmatrix}
    \end{split}  
\end{equation*}
\vspace{1em}

We next derive the permutations $(01)$, $(012)$, and $(021)$. We may perform this calculation using the properties of $S_3$, but a digression into notation is warranted. There are two conventions for evaluating a product of cycles. Recalling that cycles are permutations, which are functions, a product of cycles is thus the composition of functions, which is written in contraposition. However, within programming or engineering, one occasionally writes cycles in juxtaposition, or ``in wire,'' mimicking the order that operations appear in a circuit diagram. We use the mathematical convention for these products, aiming to provide symmetric factorizations of the necessary permutations where possible. As a convenient consequence, we may directly replace permutations with their representations by matrices to produce equivalent matrix representations. With this in mind, we see that:

\begin{equation*}
    \begin{split}
        (01)&=(02)\circ(12)\circ (02)\\
        &=CH\cdot Z_3 \cdot CH \cdot CH \cdot CH \cdot CH \cdot Z_3 \cdot CH\\
        &=CH\cdot Z^{\dagger}_3 \cdot CH\\
        \\
        \text{SHIFT+1}&=(012)\\
        &=(12)\circ(02)\\
        &=CH \cdot CH \cdot CH \cdot Z_3 \cdot CH\\
        &=CH^{\dagger }\cdot Z_3 \cdot CH\\
        \\
        \text{SHIFT-1}&=(021)\\
        &=(02)\circ (12)\\
        &=CH \cdot Z_3 \cdot CH \cdot CH \cdot CH\\
        &=CH \cdot Z_3 \cdot CH^{\dagger}
    \end{split}
\end{equation*}

Thus establishing the universality of the set \{$CH,Z_3$\} for permutative ternary quantum circuits.
\end{proof}

Our goal is to have the cost of all permutative gates as equal as possible, as this simplifies the higher level synthesis algorithms for more complex ternary circuits. Towards this end we replace the set \{$CH, Z_3$\} with the set \{$CH, CH^\dagger, Z_3, Z^\dagger_3$\}, where $CH^\dagger = CH^3$, obtaining a non-minimal universal set for permutative ternary quantum circuits. By similar methods  many other pairs of gates that are universal with respect to permutative ternary quantum circuits can be established. These pairs are given by taking a permutative variant of the Chrestenson gate and a diagonal gate appropriate to that permutative form. We further note that in place of the diagonal transformation, we can also use $SB$ or $SB^\dagger$ from Figure \ref{fig:nonqft_matrices}, with a permutative Chrestenson variant, to form a universal uniformizing gate pair for permutative ternary quantum circuits.

\subsection{Ternary Toffoli}
\begin{figure}[H]
  \centering
  \includegraphics[width=0.45\linewidth]{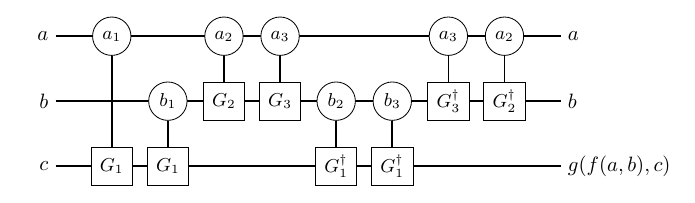}   
  \caption{A general circuit template for use in the implementation of ternary functions of the form $g(f(a,b),c)$.}
  \label{fig:three-com-gf3}
\end{figure}

For the Toffoli Galois(3) gate in quantum ternary logic, the essence of its circuit design lies in how to achieve the modulo-3 multiplication $\cdot_3$ of two control qutrits in ternary logic and the modulo-3 addition $+_3$, by the combination of two-qutrit gates (i.e., single-controlled quantum ternary unitary gates) and single-qutrit gates.
Generally, the circuit design of the two-controlled unitary gates in quantum ternary logic is also based on these two basic operations. 
Therefore, based on Barenco's symmetrical design for the binary quantum Toffoli gate in \cite{barenco1995elementary}, a similar quasi-symmetrical reversible circuit template is proposed as shown in Figure \ref{fig:three-com-gf3}.

In this circuit template diagram, $a_i$ and $b_i$ represent any pure quantum basis states corresponding to the control qutrits, that is, $a_i, b_i \in \{ | 0\rangle, |1\rangle, |2\rangle \}$. The matrices $G_i$ and $G_i^\dagger$ respectively represent any ternary quantum gate and its adjoint. As these matrices represent unitary transformations, $G_i^\dagger=G_i^{-1}$.

\begin{figure}[H]
  \centering
  \includegraphics[width=0.90\linewidth]{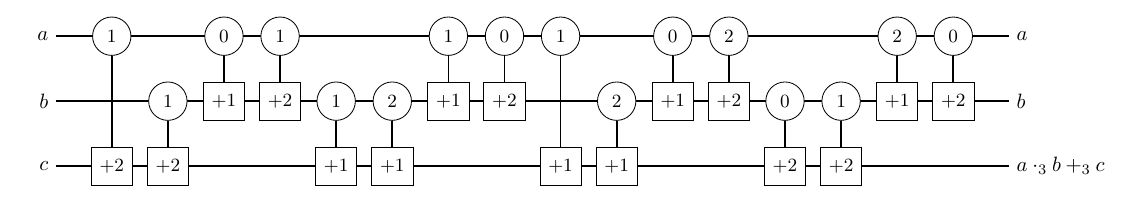}   
  \caption{The circuit is the general structure to implement the $a\cdot_3 b +_3 c$ on GF(3) for arbitrary $a$, $b$, and $c$. $a,b,c$ are all qutrits, $\cdot_3$ the multiplication here is the multiplication on GF(3). The add $+_3$ here is the addition on GF(3).}
  \label{fig:abc-gf3}
\end{figure}

As an example, we consider the function $a\cdot_3b+_3c$. This example will utilize two instances of our template in Figure~\ref{fig:three-com-gf3} and give us a circuit design for the ternary Toffoli gate which subsequently should be optimized via local transformations. This circuit design can be used to represent the execution of the same gate operation under three minimum control conditions, similar to the single-control quantum ternary unitary gate, but without the constraint that these three minimum control conditions are on the same control qutrit, and can represent the different three possible combinations of constraints of two control qutrits. 
Based on this design, by combining the circuits that respectively represent the control conditions of $\{ |00\rangle, |11\rangle, |22\rangle \}$ undergoing a SHIFT$+1$ gate operation and those representing the control conditions of $\{ |00\rangle, |12\rangle, |21\rangle \}$ undergoing a SHIFT$+2$ gate operation, the output of the function $a \cdot_3 b +_3 c$ over GF(3) can be achieved, as shown in Figure \ref{fig:abc-gf3}. Given that the function $a \cdot_3 b +_3 c$ simultaneously covers the basic addition operation (corresponding to the NOT logic required by the circuit) and the multiplication operation (corresponding to the AND logic required by the circuit) over GF(3), the circuit designs for the adder and multiplier over GF(3) can both be obtained through local simplification and transformation of the circuit shown in Figure \ref{fig:abc-gf3}.

\subsection{Ternary Swap}
\begin{figure}[H]
  \centering
         \includegraphics[width=0.25\linewidth]{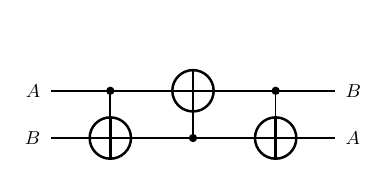} 
  \caption{Binary SWAP circuit expressed via 3 CNOTs}
  \label{fig:swapexample}
\end{figure}

The next ternary gate we consider is the ternary analogue of the binary SWAP gate. The binary SWAP gate is typically realized as the composition of three controlled not gates, as indicated in Figure~\ref{fig:swapexample}.

\begin{figure}[H]
  \centering
  \includegraphics[width=0.27\linewidth]{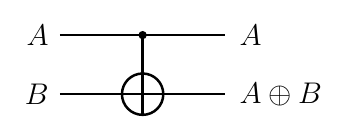} 
  \hspace{3em}
  \includegraphics[width=0.40\linewidth]{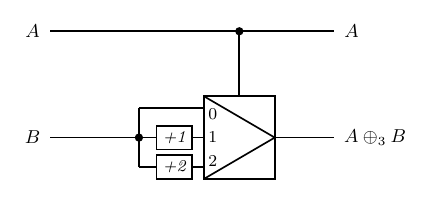}   
  \caption{Left: The binary gate for addition modulo $2$, also known as CNOT or XOR. Right: The ternary gate for addition modulo $3$, indicated symbolically by $\oplus_3$.}
  \label{fig:feynmangalois}
\end{figure}

Using this circuit as a heuristic, we can design a ternary SWAP employing ternary multiplexers in place of the binary CNOT gates, these multiplexers appear in Figure \ref{fig:feynmangalois}.

\begin{figure}[H]
  \centering
  \includegraphics{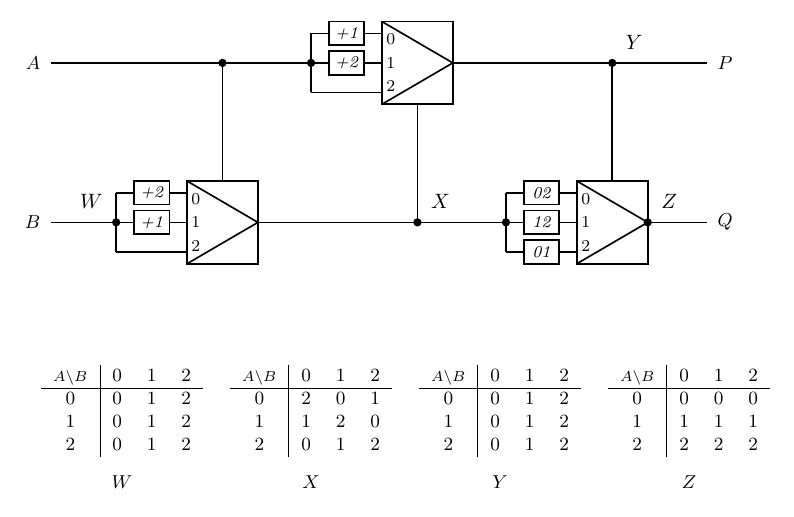}   
  \caption{A circuit for ternary SWAP built using ternary multiplexers and permutative transformations. The tables below describe the logical state at points $W$, $X$, $Y$, $Z$ as found on the top and bottom wires.}
  \label{fig:multiplexer}
\end{figure}

In Figure~\ref{fig:multiplexer}, we use tables to denote the state of the wires $A$ and $B$ at the points $W$, $X$, $Y$, and $Z$. These values vary according to the values of the original ternary inputs on $A$ and $B$. For instance, the table labeled $X$ describes the logical state of the $B$ wire at point $X$, as dependent on the inputs on the $A$ and $B$ wires. Within this table the $A$ states are indicated by the rows and the $B$ states indicated by the columns. In particular, if the original input on wire $A$ is $1$ and the original input on wire $B$ is $2$, then the state of $B$ at point $X$ is $0$, as given in the table. We see that by the end of the circuit, the table $Z$, representing the state of wire $B$ at point $Z$, has at each position the same value as the original input on wire $A$. Accordingly the logical state of wire $A$ has been swapped into wire $B$. This is a standard approach for synthesizing novel logical gates and circuits in multi-valued logic, and is an extension of the standard truth-table method in binary logic.

\begin{figure}[H]
    \centering
    \includegraphics[width=0.60\linewidth]{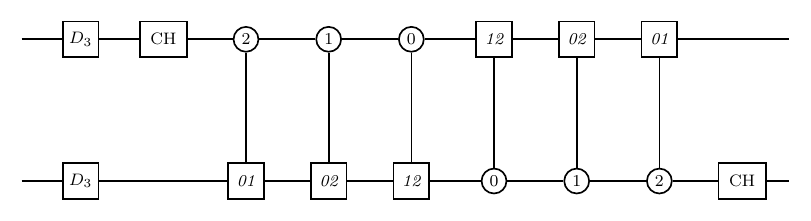} 
    \caption{}
    \label{fig:50_swap_c}
\end{figure}

Much like the binary “iSWAP” gate (a SWAP gate with phase) in the IBM library~\cite{IBM_iSwapGate_2025}, this circuit can be enhanced using the Chrestenson transformation, as shown in Figure \ref{fig:50_swap_c}.
In addition to the permutation effect, the two-qutrit output of the circuit in Figure \ref{fig:50_swap_c} acquires a phase of $\omega$ when the input is $|02\rangle$, $|20\rangle$, or $|22\rangle$. If the input is $|11\rangle$, $|12\rangle$, or $|21\rangle$, the output phase becomes $\omega^2$. For all other inputs, no phase is introduced.

\subsection{MIN / MAX}
We now turn our attention to gates which sift out the minimum and maximum logical values. Gates of this kind appear in several multivalued logics in the literature, such as the {\L}ukasiewicz and Post logics~\cite{muzio1986, dubrova1996, dubrova1999,stankovic1998, green1974}. While many authors work on Galois Field based realizations of quantum reversible circuits, there are several other possible logic systems to realize arithmetic operators, general logic and algebra concepts in quantum permutative circuits. 
There are two groups of logic:
\begin{enumerate}
    \item Logics with Min and Max operations, such as classical Łukasiewicz and Post logics~\cite{muzio1986}.
    \item Logics based on binary Reed-Muller (Zhegalkin) logic generalized to multi-valued \cite{muller1954,reed1954,zhegalkin1927}. We can call them fixed-polarity generalizations of standard canonical Reed-Muller forms and non-canonical circuits such as Exclusive-Or-Sum of Products (ESOP).
\end{enumerate}

The most important systems in the first group of logics include the following:

\begin{enumerate}
    \item {\L}ukasiewicz (1920)  introduced a ternary logic, next extended (1922) to any number n of values. {\L}ukasiewicz gates are: implication $A\implies B = \max(1-A,B)$, negation $A = 1-A$, Conjunction $A \wedge B = \min(A,B)$, Disjunction $A \vee B = \max(A,B)$. It has applications in formal logic and algebraic systems~\cite{muzio1986}.
    \item Post System (1920) is for any $n \geq 2$ and in addition to {\L}ukasiewicz operators it includes the  Biconditional defined as $A \Longleftrightarrow B =  \max ((A\implies B) , (B \implies A))$~\cite{muzio1986}.
    \item  Kleene’s Logic (K3) is a logic of indeterminacy (K3) and  Priest’s ``logic of paradox'' (P3) are both ternary. The gates include Min, Max, modified negation, and implication gates~\cite{kleene1952,Priest1979-PRITLO}.
    \item Gödel-Dummett Logics is a famil of continuous logics on interval [0,1] and includes Min, Max and Product with applications in Fuzzy Logic~\cite{Dummett1959, Godel1932-GDEZIA}.
\end{enumerate}

\begin{figure}[H]
  \centering
  \includegraphics[width=0.90\linewidth]{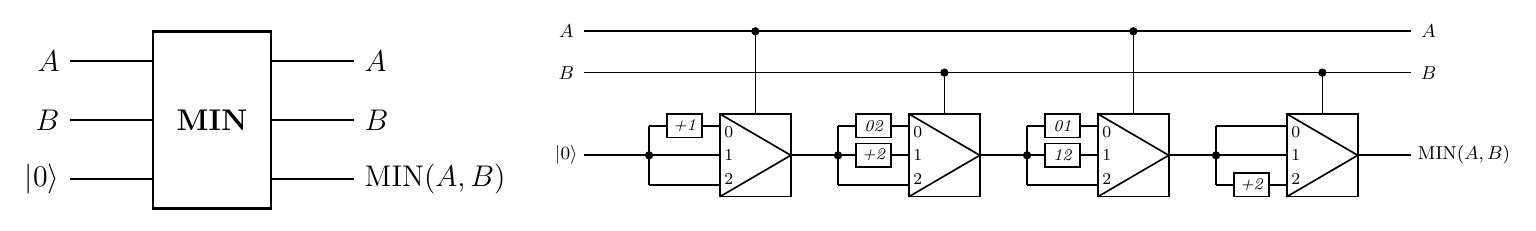}   
  \caption{A stylistic and realized circuit for the MIN operator.}
  \label{fig:mingate}
\end{figure}

In Figure~\ref{fig:mingate} a circuit with quantum multiplexers is illustrated that resulted from a computer search via exhaustive techniques to find the circuit for the MIN operator with a minimum number of quantum multiplexers.

\begin{figure}[H]
  \centering
  \includegraphics[width=0.90\linewidth]{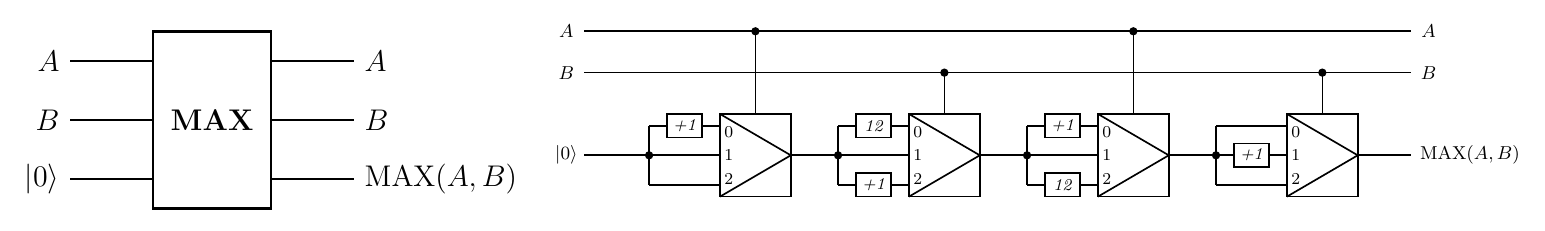}   
  \caption{A stylistic and realized circuit for the MAX operator.}
  \label{fig:maxgate}
\end{figure}

In Figure~\ref{fig:maxgate} a circuit with quantum multiplexers is illustrated that resulted from a computer search via exhaustive techniques to find the circuit for the MAX operator with a minimum number of quantum multiplexers.

Usually the process of synthesizing a quantum circuit starts from an expression in Post-like logic or Galois-like logic. Then every operator in such an expression is replaced with quantum multiplexers as illustrated in Figures \ref{fig:mingate} and \ref{fig:maxgate}. 

This process continues with the replacement of the multiplexers with the appropriate gates selected via the permutative circuit notation as discussed in Section~\ref{sec:ternaryquantumlogictransforms}. This step is followed by the replacement of every permuative gate with its synthesis in Chrestenson and $Z$ gate variants. There are many ways to achieve this circuit synthesis and the optimization process requires the finding of the most cost effective circuit. This is achieved through the use of local equivalence transformations.

\section{Future Directions}
\begin{figure}[H]
    \centering
    \includegraphics[width=0.75\linewidth]{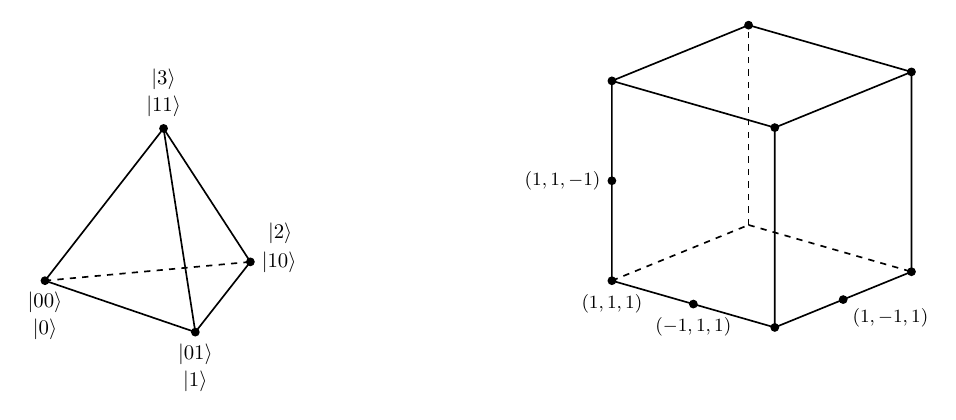}
    \caption{The toric variety visualization of the pure state space of a four-state quantum system. The vertices of the tetrahedron containing the convex coordinates are labeled with the states of a two-qubit register and the states of a single ququadit. Three uniform states over the barycenter are illustrated on the right, indicating which axis corresponds to which projective coordinate via the right hand rule.}
    \label{fig:cp3_vis}
\end{figure}

\begin{figure}[H]
    \centering
    \includegraphics[width=0.75\linewidth]{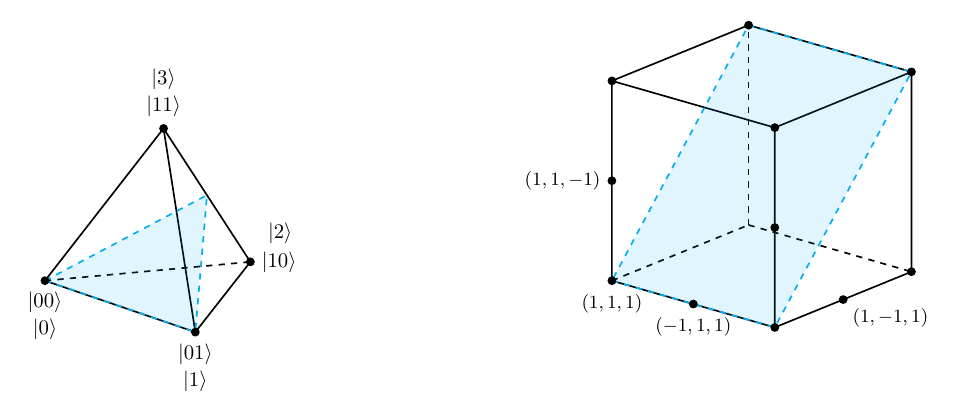}
    \caption{A visualization of the CNOT transformation on a register of 2 qubits, equivalent to the radix-4 transposition (23). These transformations induce separate reflections in the convex coordinates and in the periodic coordinates.}
    \label{fig:cp3_cnot}
\end{figure}

Our techniques extend to the quaternary and two-qubit register case, which is the subject of a later publication. It is interesting to note that the state space for radix-4 quantum computation and the state space for a register of two qubits are identical---complex projective three space.  Accordingly, the same visualization may be used for both, as pictured in Figure~\ref{fig:cp3_vis}. Further, every two-qubit quantum transformation can be considered as a single ququadit transformation, however, the converse is not in general true since $\text{SU(2)}\otimes\text{SU(2)}$ is a proper subgroup of $SU(4)$ \cite{draayer1970}. Transformations in SU(4) and those in $\text{SU(2)}\otimes\text{SU(2)}$ can be visualized in our model, allowing for the visualization of singly controlled two-qubit transformations such as CNOT (see Figure~\ref{fig:cp3_cnot}).

A further application of this visualization may be found in the physics literature \cite{bengtsson2002}, which has employed the toric variety structure of $\CP^3$ to visualize the separable and maximally entangled states of a register of two qubits.

Future tasks include
working with basic gate sets as realized in various industrial implementations, and use of the visualization techniques developed herein to create algorithms to determine minimal cost gates and optimal quantum circuit synthesis methods in ternary, as performed for binary quantum computation in \cite{Hung2004QuantumLogicSynthesis,Hung2006OptimalSynthesis}. Additional tasks include the creation of a library of optimal circuits for the fundamental optimal multivalued circuits for the fundamental transformations such as Toffoli and SWAP, in addition to optimized multivalued circuits for arbitrary functions of 3 and 4 variables. Yet another task is the visualization of the transformations for mixed registers of quantum dits (e.g. a register consisting of a single qubit and a single qutrit) with the goal of obtaining optimized circuits for mixed registers. Another is development of expressions for the circuits we have designed here, in particular the extension of the existing research of Post, {\L}ukasiewicz, et al. Similarly, develop a set of local optimization transformations analogous to those currently existing for binary quantum circuits.

Finally, here is a specific open problem to be addressed. Assuming costs for a certain selection of basic gates, create a provable exact minimal cost circuit. In particular, suppose the cost  of $CH$ is 2, the cost of $Z$ is 1, and the cost $CZ$ is 2. For the ternary adder, find the circuit with exact provable minimal cost.

\section{Summary}
In this paper, we pointed out the concurrence of toric geometry and quantum mechanical structures on the state spaces of quantum computational units, in particular the identification of the equivalence classes of quantum states under measurement with the orbits of the toric geometric structure of the finite dimensional complex projective spaces. We provided visualizations of these state spaces and of certain fundamental transformations in binary and ternary quantum logic and a method to develop new transformations based on these visualization techniques. Transformations discussed included minimal universal sets for permutative ternary quantum circuits. In addition, general structures and synthesis methods based on quantum multiplexers were presented. Also presented was a general framework for the design of optimal ternary quantum transformations and circuits. Finally, a number of open research areas that are extensions of the work presented herein were given.

\backmatter

\bmhead{Acknowledgements}
Not applicable.

\section*{Declarations}

\begin{itemize}
  \item \textbf{Funding}: This research received no external funding.
  \item \textbf{Conflict of interest}: On behalf of all authors, the corresponding author states that there is no conflict of interest. All authors have read and agreed to the published version of the manuscript.
  \item \textbf{Ethics approval and consent to participate}: Not applicable.
  \item \textbf{Consent for publication}: Not applicable.
  \item \textbf{Data availability}: The original contributions presented in our study are included in this article; further inquiries can
be directed to the corresponding author (B.).
  \item \textbf{Materials availability}: Not applicable.
  \item \textbf{Code availability}: Search code available upon request.
  \item \textbf{Authors' contribution}:  Visualization, B.; 
  Conceptualization B. A. C. P.; 
  Methodology, B., A., P., C.;
  Formal analysis B., A. , C. P.;
  Writing B., A., P., C., W., O., E.R., J.R.;
  Document preparation, W., O., E.R., J.R.;
  Development and verification of formulae W., O., E.R., J.R.;
  Literature search and reference verification, W., O., E.R., J.R.;
  Coordination P.;
  Circuit Synthesis C., J., C..

\end{itemize}

\bibliography{citations}

\end{document}